\documentclass[12pt]{article}
\usepackage{amsmath}
\usepackage{amssymb}
\usepackage{graphicx,psfrag,epsf}
\usepackage{enumerate}
\usepackage{natbib}
\usepackage{multirow}
\usepackage{array, booktabs}
\usepackage{lscape}
\usepackage[ruled,linesnumbered]{algorithm2e}
\usepackage{hyperref}
\newcommand{\blind}{0}

\makeatletter
\newcommand{\removelatexerror}{\let\@latex@error\@gobble}
\makeatother

\begin{document}

\def\spacingset#1{\renewcommand{\baselinestretch}%
{#1}\small\normalsize} \spacingset{1}

\newcolumntype{P}[1]{>{\raggedright\arraybackslash}p{#1}}
\newcommand{\ra}[1]{\renewcommand{\arraystretch}{#1}}


\if0\blind
{
  \title{\bf Detecting and Classifying Outliers in Big Functional Data}
  \author{Oluwasegun Taiwo Ojo
    \\
    IMDEA Networks Institute\\
    Universidad Carlos III de Madrid \\ 
    Antonio  Fern\'andez Anta \\
    IMDEA Networks Institute \\
    Rosa E. Lillo  \\
    uc3m-Santander Big Data Institute\\
Department of Statistics,\\
Universidad Carlos III de Madrid \\ 
    Carlo Sguera \\
    uc3m-Santander Big Data Institute \\
    Universidad Carlos III de Madrid
    }
  \maketitle
} \fi

\if1\blind
{
  \bigskip
  \bigskip
  \bigskip
  \begin{center}
    {\LARGE\bf Detecting and Classifying Outliers in Big Functional Data}
\end{center}
  \medskip
} \fi

\bigskip
\begin{abstract}
We propose two new outlier detection methods, for identifying and classifying different types of outliers in (big) functional data sets. The proposed methods are based on an existing method called Massive Unsupervised Outlier Detection (MUOD). MUOD detects and classifies outliers by computing for each curve, three indices, all based on the concept of linear regression and correlation, which measure outlyingness in terms of shape, magnitude and amplitude, relative to the other curves in the data. `Semifast-MUOD', the first method, uses a sample of the observations in computing the indices, while `Fast-MUOD', the second method, uses the point-wise or $L_1$ median in computing the indices. The classical boxplot is used to separate the indices of the outliers from those of the typical observations. Performance evaluation of the proposed methods using simulated data show significant improvements compared to MUOD, both in outlier detection and  computational time. We show that Fast-MUOD is especially well suited to handling big and dense functional datasets with very small computational time compared to other methods. Further comparisons with some recent outlier detection methods for functional data also show superior or comparable outlier detection accuracy of the proposed methods. We apply the proposed methods on weather, population growth, and video data.

\end{abstract}

\noindent%

\spacingset{1.45}
\section{Introduction}
\label{sec:intro}

Technological advances in the latest decades have allowed the observation of data samples that can be considered as functions or curves over a domain. These include temperature data over time, pixel values of images, frames of video, etc., and it is natural to assume that these observations have been generated by a stochastic function over a domain. Functional data analysis (FDA) deals with statistical analysis of these types of data. We refer the reader to \cite{fdaramsay2005} for an overview of statistical methods for analysing functional data. Non-parametric methods for FDA have also been treated in \cite{ferraty2006nonparametric}, while a survey of theory of statistics for FDA can be found in \cite{cuevas2014partial}. 


It is common practise to identify outliers before conducting statistical analyses. Outliers are of interest because they could significantly bias the results of statistical inference. Furthermore, an outlier, rather than being due to a measurement error, could be due to some interesting changes or behaviour in the data-generating process, and it is often of interest to investigate such changes. This is even more important in the analysis of weather, pollution, and geochemical data where identifying such changes is necessary to make important environmental policy decisions (e.g., \cite{filzmoser2005geochem}). In the context of FDA, identifying outliers becomes even more difficult because of the nature of functional observations. Such observations are realizations of functions over an interval and thus, outlying observations could have extreme values in a part of the interval or in all the interval. These (outlying) functional observations could exhibit different properties which make them anomalous. These include being significantly shifted from the rest of the data or having a shape that on the average is different from the rest of the data. \cite{hubert2015multivariate} defined the former as magnitude outliers, the latter as shape outliers, and in addition defined amplitude outliers as curves or functions which may have the same shape as the mass of the data but with different amplitude. 

Outliers in multivariate data are typically identified using notions of statistical depth, which provide a centre-outward ordering for observations. Statistical depths were generalized to the functional domain starting with the work of \cite{fraiman2001trimmed}. Since then, various depth notions for ordering functional data have been introduced, including band depth and modified band depth (\citeauthor{Romo_depths} \citeyear{Romo_depths}), extremal depth (\citeauthor{narisetty2016extremal} \citeyear{narisetty2016extremal}),  half-region depth (\citeauthor{lopez2011half} \citeyear{lopez2011half}), and total variation depth (\citeauthor{huang2019decomposition} \citeyear{huang2019decomposition}), among others (see \cite{nieto2016topologically} for more details). A number of exploratory and outlier detection methods for functional data are based on functional depth notions. For instance, \cite{febrero2008} proposed an outlier detection method using functional depths with cutoffs determined through a bootstrap, while \cite{sguera2016functional} proposed to use a kernelized functional spatial local depth (KFSD) for identifying outliers. The functional boxplot (\citeauthor{sun2011functional} \citeyear{sun2011functional}) also uses the modified band depth to define a 50\% central region on a functional data with curves outside 1.5 times the central region flagged as outliers, analogous to the classical boxplot. Likewise \cite{arribas2014shape} proposed the outliergram, which uses the quadratic relationship between the modified band depth and the modified epigraph index (\citeauthor{lopez2011half} \citeyear{lopez2011half}) to identify shape outliers. Other methods, like the functional bagplot or the functional highest density regions (\citeauthor{hyndman2010rainbow} \citeyear{hyndman2010rainbow}), use the first two principal components of the functional data to construct a bagplot or a highest density region plot, respectively, to identify outliers. \cite{hubert2015multivariate} also proposed using a bag distance and skewness adjusted projection depth to identify outliers. 

More recent literature include the work of \cite{dai2018multivariate}, in which they constructed a magnitude-shape plot (MS-Plot) for visualizing the centrality of multivariate functional observations and for identifying outliers, using a functional directional outlyingness measure. This functional directional outlyingness measure for multivariate functional data was further investigated in \cite{dai2019directional}. Furthermore, \cite{rousseeuw2018measure} introduced another measure of functional directional outlyingness for multivariate functional data and used it to construct the functional outlier map (FOM) for identifying outliers in multivariate functional data, while \cite{huang2019decomposition} defined the shape similarity index and the modified shape similarity index based on total variation depth to identify shape outliers.  \cite{dai2020sequential} proposed to use some predefined sequence of transformations to identify and classify shape and amplitude outliers after first removing the magnitude outliers using a functional boxplot. 

It is desirable to be able to identify all the different types of outliers in functional data. However, some outlier detection methods for functional data are specialized, in the sense that they are well suited to identifying outliers of a certain type; e.g., outliergram is well suited to identifying shape outliers, while functional boxplot is well suited to identifying magnitude outliers. While some methods are sensitive to different types of outliers, they do not automatically provide information on the type of outliers, unless the data is visualized. Thus, it might be difficult to understand why a particular curve is flagged as an outlier. This is especially important when the functional data is large and not easy to visualize.  Classifying the types of outliers also allows for selectively targeting different types of outliers. For example, one might be interested only in shape outliers or only in magnitude outliers. Furthermore, some methods do not scale up to large functional datasets, which poses a challenge with the huge amounts of data that is being generated nowadays. 

In this article, we introduce two new outlier detection methods for univariate functional data: Fast-MUOD and Semifast-MUOD, which are based on the Massive Unsupervised Outlier Detection (MUOD) method proposed in \cite{azcorra2018unsupervised}. These methods are capable of identifying and classifying magnitude, amplitude and shape outliers without the need for visualization. The proposed methods are based on the concepts of linear regression and correlation, making them quite intuitive and easy to compute. We also show that one of the proposed methods, Fast-MUOD, scales quite well and is thus suitable for detecting outliers in big functional data. We show that these methods have good outlier detection performance on a range of outlier types using simulation experiments and we also compare positively, their outlier detection performance and computation time to some existing outlier detection methods for functional data. The main contributions of this work are: 
\begin{itemize}
    \item[-] Proposal of two new methods capable of identifying and classifying outliers in  functional data.
    \item[-] Simulation study comparing the proposed methods and some other recent outlier detection methods.
    \item[-] Time benchmark (comparing the proposed methods and other outlier detection methods) showing the computational time of the proposed methods.
    \item[-] Case studies showing how the proposed methods can be used in a real application and comparisons with some existing work with similar case studies.
    \item[-] An implementation of the proposed methods available on Github. 
\end{itemize}

The rest of the article is organized as follows: Section \ref{sec:meth} provides an overview of MUOD. In Section \ref{sec:muod_improv}, we present the proposed methods. These are followed by performance evaluation with some simulation studies in Section \ref{sec:simstudy}. We illustrate in Section \ref{sec::app}, the use of the proposed methods on a variety of real datasets and use cases, including object detection in surveillance video, outlier detection in weather data, and discovering growth trends in population data. We end the article with some discussions and conclusions in Section \ref{sec:conc}. 

\section{The MUOD Method}
\label{sec:meth}
In this section, we present a brief primer on MUOD as described in the supplementary material of \cite{azcorra2018unsupervised}. MUOD identifies outliers by computing for each observation or curve, three indices, namely shape, magnitude and amplitude indices. These indices measure how outlying each observation is as regards its shape, magnitude and amplitude, compared to the other observations. The definition of these indices as defined in \cite{azcorra2018unsupervised} is introduced in the following.

Consider a set of functional observations $\{Y_i\}_{i = 1}^n \in \mathcal{C(I)}$, defined on $d$ equidistant points of an interval $\mathcal{I} \in \mathbb{R}$, where $\mathcal{C(I)}$ is the space of real continuous functions defined on $\mathcal{I}$. We assume that $Y_i$ follows a distribution $F_Y$ also defined on $\mathcal{C(I)}$.  We define the MUOD shape index of $Y_i$ with respect to $F_Y$, denoted by $I_S(Y_i, F_Y)$, as

\begin{equation}
I_S(Y_i, F_Y) = \left|\frac{1}{n}\sum\limits_{j=1}^n \hat{\rho}(Y_i, Y_j) - 1 \right|,
\label{eqn::shape_muod_index}
\end{equation}
where $\hat{\rho}(Y_i, Y_j)$ is the estimated Pearson correlation coefficient between $Y_i$ and $Y_j$, given by 

$$\hat{\rho}(Y_i, Y_j) = \frac{\text{cov}(Y_i, Y_j)}{s_{Y_i}s_{Y_j} },\ \ \ \ \ \ s_{Y_i}, s_{Y_j} \ne 0$$




The correlation coefficient is responsible for capturing the similarity between each pair of curves ($Y_r$, $Y_s$) in terms of shape. The intuition behind the MUOD shape index is as follows. Assume that the number of outlying curves $n_o$ is much less than the number of non-outlying curves $n_n$, i.e., $(n_o << n_n)$. Let $Y_i$ be a normal curve (in terms of shape) with respect to (w.r.t.) $F_Y$ and let $Y_k$ be a shape outlier w.r.t. $F_Y$. Also, denote by $\{Y_j\}_{j = 1}^{n_n}$, the set of normal curves and by $\{Y_l\}_{l = 1}^{n_o}$ the set of outlying curves. Since $Y_i$ has a similar shape w.r.t. $F_Y$ and $Y_k$ is a shape outlier, the correlations between the pairs $(Y_i, Y_j)_{j = 1}^{n_n}$ will be close to $1$ and greater than the correlations between the pairs $(Y_k, Y_j)_{j = 1}^{n_n}$. Also, the correlations between the pairs $(Y_i, Y_l)_{l = 1}^{n_o}$ and $(Y_k, Y_l)_{l = 1}^{n_o}$ could be any value between $-1$ and $1$. However, $n_o << n_n$ ensures that the average of the correlations over all possible pairs $(Y_i, Y_m)_{m = 1}^n$ is greater than the average of the correlation overall all possible pairs $(Y_k, Y_m)_{m = 1}^n$. Consequently, subtracting these averages from $1$ assigns $Y_i$ a lesser shape index compared to the shape index of $Y_k$. We illustrate this behaviour of the MUOD shape index below. We generate $99$ non-outlying curves from the model
\begin{equation}
Y(t) = a_1\sin(t) + a_2 \cos(t), 
\label{eqn::mass_illustration}
\end{equation}
where $t \in {T}$, with $T$ made up of $d = 50$  equidistant domain points between $0$ and $2\pi$, and both $a_1$ and $a_2$ generated from independent uniform random variables between $0.75$ and $1.25$. Moreover, we generate a single shape outlier using the following different model:

\begin{figure}[htbp!]
	\centering
	\includegraphics[width=.7\linewidth]{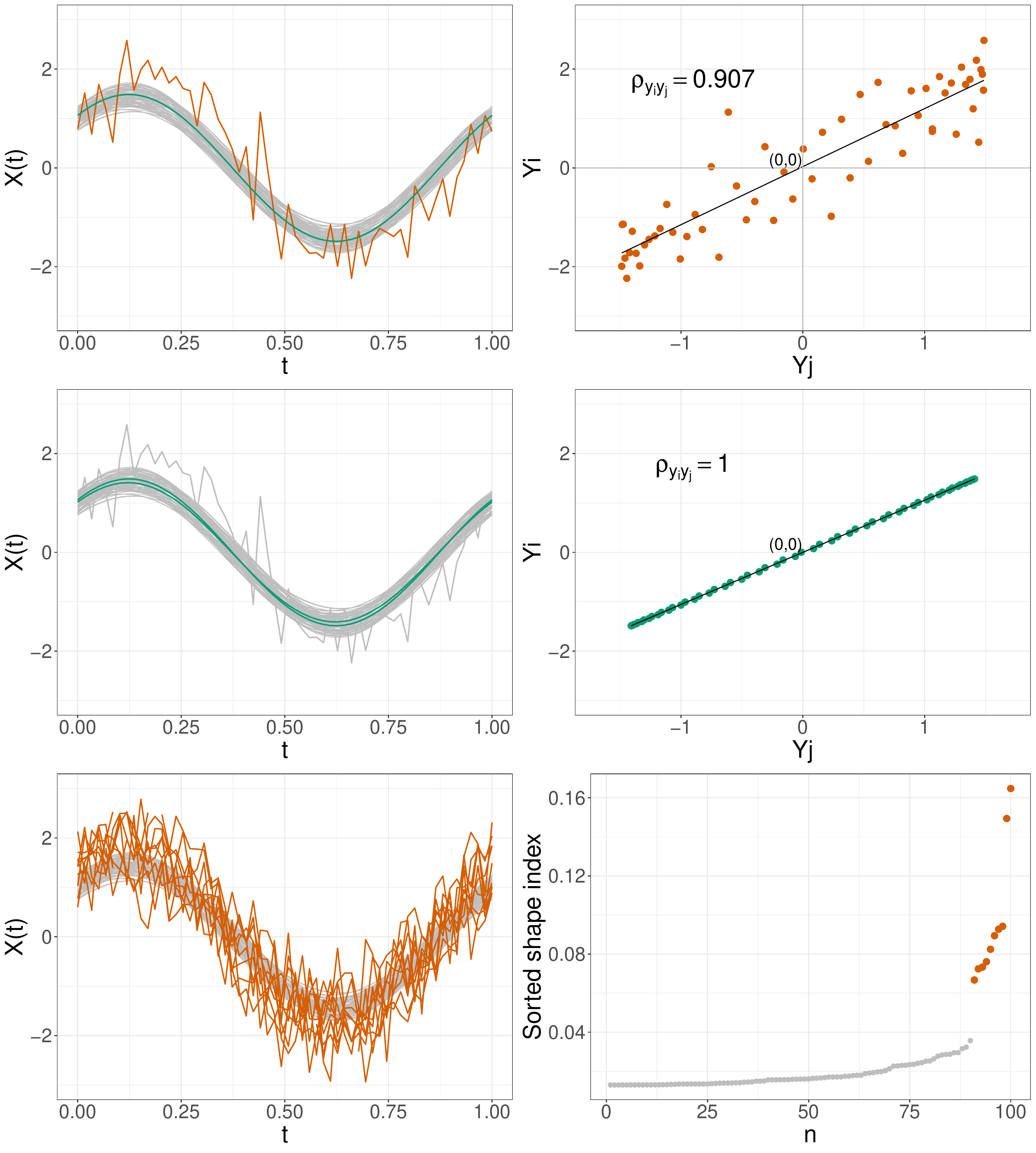}
	\caption{First Row Left: simulated data using Equation \eqref{eqn::mass_illustration} (99 curves, 98 in gray, 1 in green) and Equation \eqref{eqn::shape_illustration} (1 curve, in orange). First Row Right: estimated correlation coefficient between the observed points of the orange curve and the green curve. Second Row Left: same as First Row Left, highlighting two normal curves (green). Second Row Right: estimated correlation coefficient between the green curves. Third Low Left: Simulated data set using Equation \eqref{eqn::mass_illustration} for normal curves (in gray) and Equation \eqref{eqn::shape_illustration} for outliers (orange). Third Row Right: associated sorted MUOD shape indices.\label{fig:shape_illustration}}
\end{figure}

\begin{equation}
Y(t) = b_1\sin(t) + b_2 \cos(t)+ \epsilon(t),
\label{eqn::shape_illustration}
\end{equation}
where each $\epsilon(t)$ is drawn from a normal random variable with $\mu = 0$ and $\sigma^2 = 1/4$, and both $b_1$ and $b_2$ are realizations of independent uniform random variables with parameters $0.75$ and $1.75$. In the first row of Figure \ref{fig:shape_illustration}, we highlight a typical observation in green and the single shape outlier in orange (left) and show their estimated correlation coefficient (right). In the second row of Figure \ref{fig:shape_illustration}, we show the estimated correlation coefficient for two typical curves colored in green. We observe that the estimated correlation coefficients of the two typical curves is greater than that of the ``typical-outlier" pair of curves. In the third row of Figure \ref{fig:shape_illustration}, we show the MUOD shape indices of $90$ typical curves and $10$ outliers generated from Equations \ref{eqn::mass_illustration} and \ref{eqn::shape_illustration} respectively. Clearly, the indices of the shape outliers (in orange) are greater than the indices of the typical observations (in grey). We note that the conditions $s_{Y_i} \ne 0$  and $s_{Y_j} \ne 0$ in Equation \ref{eqn::shape_muod_index} can easily be broken if any of the curves $Y_i$ or $Y_j$ is a straight line. To avoid this, we ignore any curve $Y_j$ in the data set with $s_{Y_y} = 0$ when computing the indices.


The magnitude and amplitude indices of an observation $Y_i$, denoted by $I_M(Y_i, F_Y)$ and $I_A(Y_i, F_Y)$ respectively, are based on the intercept and slope of a linear regression between the observed points of all possible pairs $(Y_i, Y_j)_{j = 1}^n$. Let $\hat{\alpha}_j$ and $\hat{\beta}_j$ be the estimated coefficients (intercept and slope respectively) of the linear regression between  the pair $(Y_i, Y_j)$ with the observed points of the function $Y_j$ being the independent variable, and the observed points of the function $Y_i$ being the dependent variable. Then the magnitude index $I_M(Y_i, F_Y)$ of $Y_i$ is defined as:
\begin{equation}
I_M(Y_i, F_Y) = \left|\frac{1}{n}\sum\limits_{j = 1}^n\hat{\alpha}_j \right|,
\end{equation}
and the amplitude index $I_A(Y_i, F_Y)$ of $Y_i$ is defined as:
\begin{equation}
I_A(Y_i, F_Y) = \left|\frac{1}{n}\sum\limits_{j = 1}^n\hat{\beta}_j -1 \right|,
\end{equation}
with 
$$\hat{\beta}_j = \frac{\text{cov}(Y_i, Y_j)}{s_{Y_j}^2}, \ \ \ \ \ \ s_{Y_j}^2 \ne 0, 
$$
and $$\hat{\alpha}_j = \bar{x_i} - \hat{\beta}_j \bar{x}_j,$$
where 
$$\bar{x_i} = \frac{\sum_{t \in \mathcal{I}}Y_i(t)}{d}.$$
The intuition behind the magnitude index is similar to that of the shape index. We adapt the same notation used before shape outliers to magnitude outliers.  If $Y_k$ is a magnitude outlier w.r.t.$\ F_Y$, and $Y_i$ is a typical curve (in terms of magnitude) w.r.t. $F_Y$, then a linear regression between the $d$ observed points of $Y_k$ on those  of any $Y_j$ (the typical curves) will produce a large estimated intercept coefficient $\hat{\alpha}_{kj}$ compared to the esimated intercept $\hat{\alpha}_{ij}$ of the linear regression of $Y_i$ on $Y_{j}$ (since both $Y_i$ and $Y_j$ are not magnitude outliers). Provided that $n_o << n_n$, the average of the estimated $\hat{\alpha}_{km}$ values over all possible pairs $(Y_k, Y_m)_{m = 1}^n$ will be greater than the average of the estimated $\hat{\alpha}_{im}$ values over all possible pairs $(Y_i, Y_m)_{m = 1}^n$, which consequently assigns a larger magnitude index to $Y_k$, the magnitude outlier. To illustrate the magnitude index, we generate $99$ observations using Equation \eqref{eqn::mass_illustration} and a single magnitude outlier from the model below:


\begin{equation}
Y(t) = a_1\sin(t) + a_2 \cos(t) + 1.
\label{eqn::mag_out_illus}
\end{equation}
\begin{figure}[htbp!]
	\centering
	\includegraphics[width=.7\linewidth]{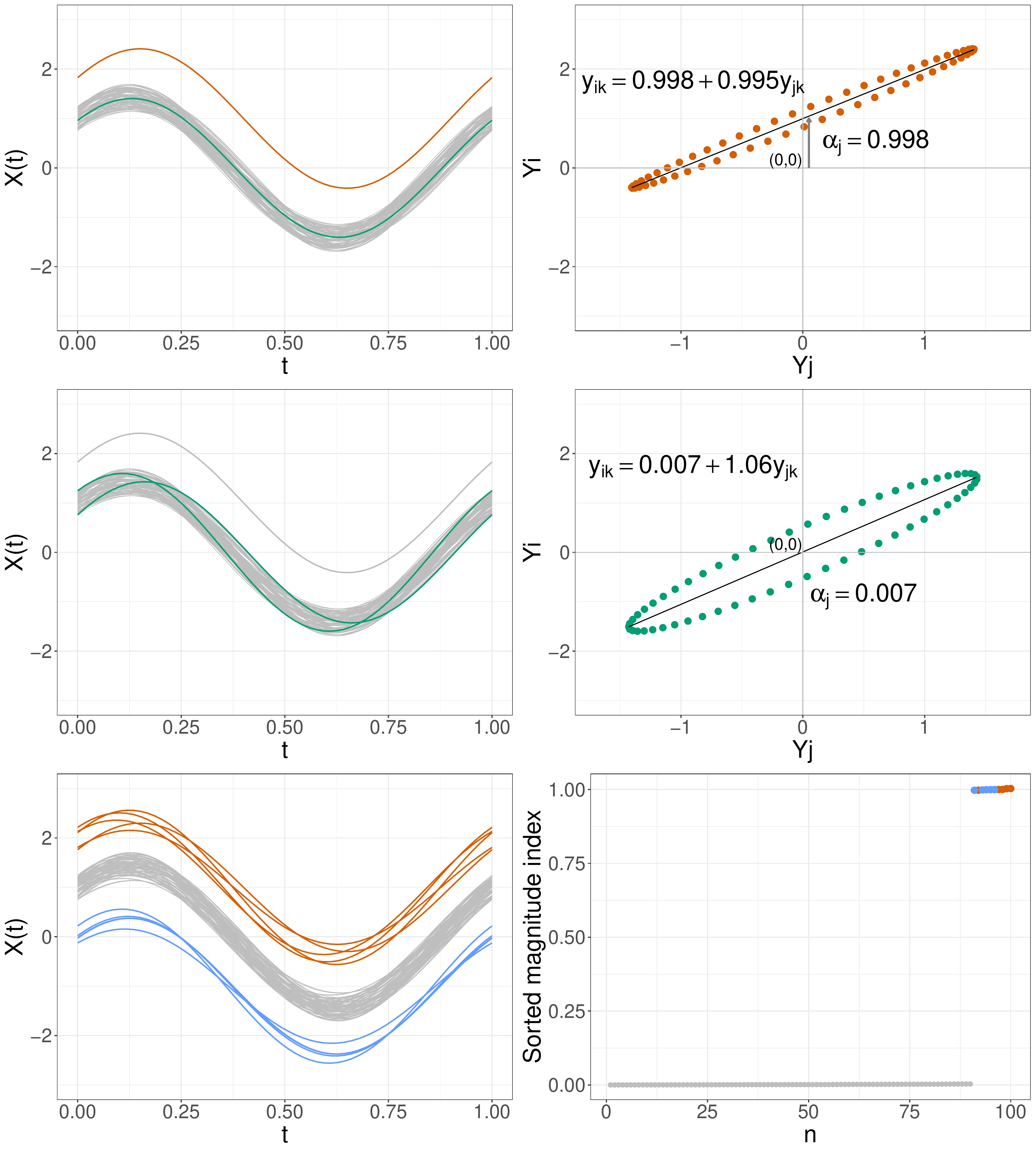}
	\caption{First Row Left: simulated data using Equation \eqref{eqn::mass_illustration} (99 curves, 98 in gray, 1 in green) and Equation \eqref{eqn::mag_out_illus2} (1 curve, in orange). First Row Right: estimated linear regression model of the orange curve on the green curve. Second Row Left: same as First Row Left, highlighting two normal curves (green). Second Row Right: estimated linear regression model between the green curves. Third Low Left: Simulated data set using Equation \eqref{eqn::mass_illustration} for normal curves (in gray) and Equation \eqref{eqn::mag_out_illus2} for outliers (in blue and orange). Third Row Right: associated sorted MUOD magnitude indices.\label{fig:mag_illus_t3}}
\end{figure}
In the first row of Figure \ref{fig:mag_illus_t3}, we show the simulated data set (left) and the estimated linear regression model between a randomly selected non-outlying curve and the unique (magnitude) outlying function together with the value of their estimated intercept (right). In the second row of Figure \ref{fig:mag_illus_t3}, we show the same simulated data set (left) and the estimated linear regression model between two randomly selected non-outlying curves (right). A comparison of the estimated intercepts (of the former and the latter pairs of functions) shows that the estimated intercept for ``normal-outlier'' pair of curves is greater than that of the ``normal-normal'' pair of curves.  Finally, in the third row of Figure \ref{fig:mag_illus_t3}, we show another simulated data set (left) where normal observations are generated using Equation \eqref{eqn::mass_illustration}, and $10$ magnitude outliers are generated using Equation \eqref{eqn::mag_out_illus2}:

\begin{equation}
Y(t) = a_1\sin(t) + a_2 \cos(t) + k,
\label{eqn::mag_out_illus2}
\end{equation}
where $k$ takes either $-1$ or $1$ with equal probability, and it controls whether an outlier is higher or lower in magnitude than the typical observations. On the right of the third row of Figure \ref{fig:mag_illus_t3}, we show the sorted MUOD magnitude indices. All the low and high magnitude outliers (in blue and orange respectively) have significantly larger indices than the typical observations.

\begin{figure*}
	\centering
	\includegraphics[width=.7\linewidth]{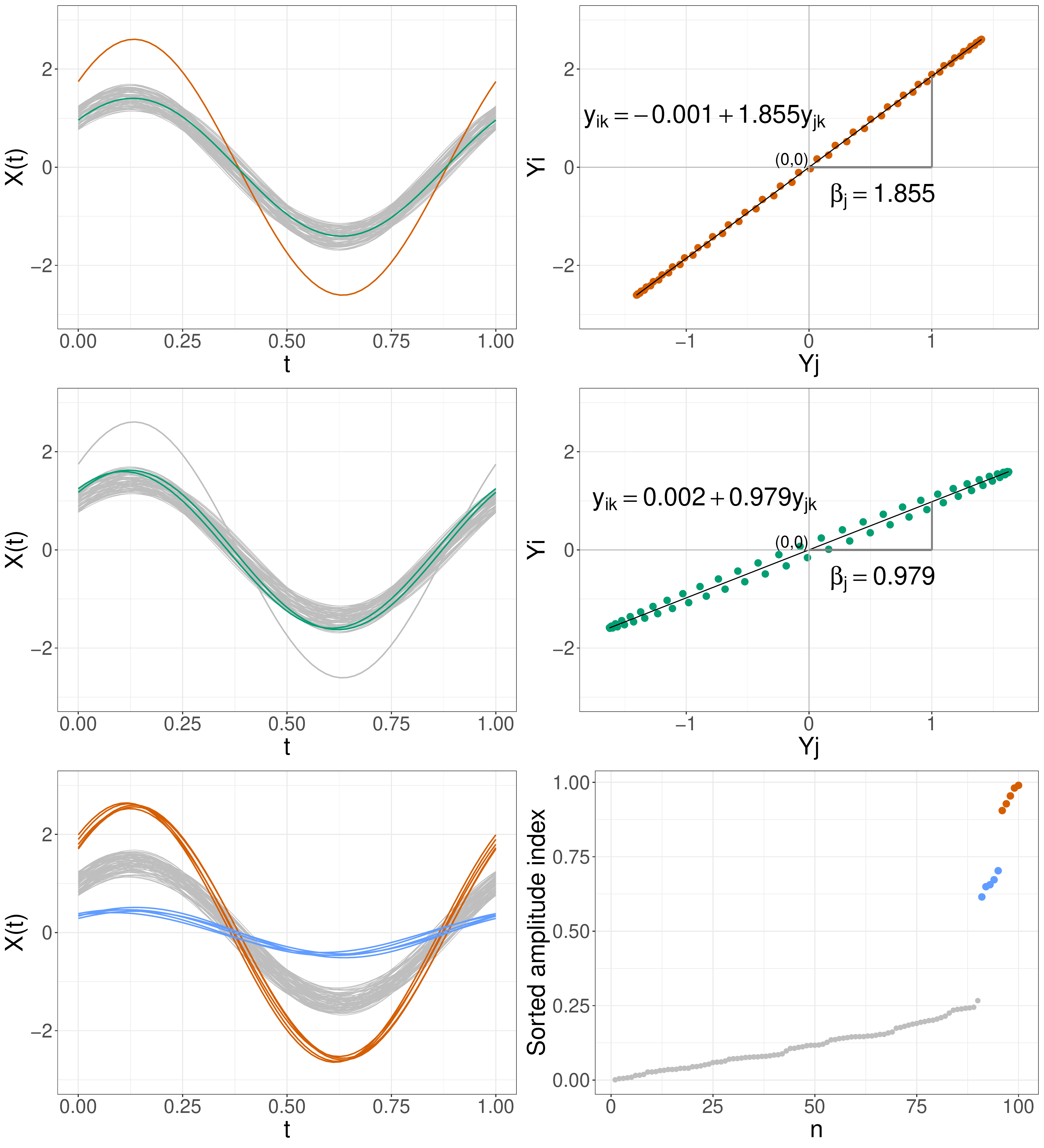}
	\caption{First Row Left: simulated data using Equation \eqref{eqn::mass_illustration} (99 curves, 98 in gray, 1 in green) and Equation \eqref{eqn::amp_out_illus} (1 curve, in orange). First Row Right: estimated linear regression model of the orange curve on the green curve. Second Row Left: as First Row Left, highlighting two normal curves in green. Second Row Right: estimated linear regression model between the green curves. Third Row Left: Simulated data set using Equation \eqref{eqn::mass_illustration} for normal curves (in gray) and Equation \eqref{eqn::amp_out_illus} for outliers (in blue and orange). Third Row Right: associated sorted MUOD amplitude indices.\label{fig:amp_illus_t3}}
\end{figure*}

Unlike the magnitude index which uses the intercept term, the amplitude index uses the slope term. The same intuition applies for the amplitude index because if both $Y_i$ and $Y_j$ are similar curves (in amplitude), increasing and decreasing in amplitude at a similar rate, then the linear regression between their $d$ observed points will produce an estimated slope coefficient $\hat{\beta}_j$ close to 1. We illustrate the amplitude index in Figure \ref{fig:amp_illus_t3}. This figure resembles Figure \ref{fig:mag_illus_t3}, but with amplitude outliers, which we generate using the model in Equation \ref{eqn::amp_out_illus}:


\begin{equation}
Y(t) = c_1\sin(t) + c_2 \cos(t),
\label{eqn::amp_out_illus}
\end{equation}
where $c_1$ and $c_2$ are independent uniform random variables between $1.7$ and $2.0$ for higher amplitude outliers; and between $0.2$ and $0.4$ for lower amplitude outliers.
From Figure \ref{fig:amp_illus_t3}, the estimated slope coefficient between the amplitude outlier (in orange) and the typical observation (in green) is $\hat{\beta}_j = 1.855$ (top row), while the estimated slope coefficient between the two typical observations is $\hat{\beta}_j = 0.979$ (second row). Moreover, the sorted MUOD amplitude indices of the amplitude outliers are greater than those of the typical observations (third row). 

After obtaining the MUOD indices as defined above, the next step in outlier detection is to differentiate the indices of the outliers from the indices of the typical observations. \cite{azcorra2018unsupervised} proposed two heuristic methods to perform this task. The first involves approximating the sorted indices with a curve and searching for a cutoff point on the curve where the first derivative of such point fulfills a certain condition (e.g., a point on the curve with first derivative greater than 2). The other method, named `\textit{tangent method}', searches for the line tangent to the maximum index and then uses as threshold the point at which the tangent intercepts the  $x$-axis. These methods are particularly prone to detecting normal observations as outliers (\citeauthor{vinue2020robust} \citeyear{vinue2020robust}). Furthermore, there is no statistical motivation behind these two proposed heuristic methods since they were mainly used as a quick support for identifying outliers in the real data application in \cite{azcorra2018unsupervised}. As part of our proposed improvements, we use a classical boxplot for separating the indices of the outliers from those of the typical curves. 

\section{Fast-MUOD and Semifast-MUOD}
\label{sec:muod_improv}
We discuss the proposed methods based on MUOD in this section. First we describe how Semifast-MUOD and Fast-MUOD compute their outlier indices. Then we present the use of the classical boxplot for identifying a cutoff for the indices. We also describe their implementations. 

\subsection{Semifast-MUOD}
\label{subsec:sfmuod}
Due to the way MUOD indices are defined, MUOD is computationally intensive and by design the time complexity for MUOD to compute its three indices is in the order of $\Theta(n^2d)$. This is because the three indices of each of the $n$ functional observations are computed by using all the $n$ observations in the data. To reduce computational time, we propose to use a sample of the observations in the computation of the three indices. We pick a random sample (without replacement) of size $n_X$, from the set of observations $\{Y_i\}_{i = 1}^n$  based on an appropriate sample proportion $p \in (0,1]$. Denote this random sample by $\{X_i\}_{i = 1}^{n_X}$ and its empirical distribution by $F_X$. Then, for each observation $Y_i$, the three indices for $Y_i$ are computed using the $n_X$ observations in $\{X_i\}_{i = 1}^{n_X}$ rather than the $n$ observations of $Y$. Formally, we define the shape index of any  $Y_i$, now with respect to $F_X$, denoted by $I_{S}(Y_i, F_X)$ as 

\begin{equation}
I_S(Y_i, F_X) = \left|\frac{1}{n_X}\sum\limits_{j = 1}^{n_X}\hat{\rho}(Y_i, X_j) - 1 \right|,
\end{equation}
where $\hat{\rho}(Y_i, X_j)$ still remains the estimated Pearson correlation coefficient between $Y_i$ and $X_j$ for $i = 1. \ldots, n$ and $j = 1, \ldots, n_X$. Likewise, we define the new magnitude and amplitude indices, $I_M(Y_i, F_X)$ and $I_A(Y_i, F_X)$, computed w.r.t. $F_X$ as

\begin{align}
I_M(Y_i, F_X) &= \left|\frac{1}{n_X}\sum\limits_{j = 1}^{n_X}\hat{\alpha}_j \right|,\\
I_A(Y_i, F_X) &= \left|\frac{1}{n_X}\sum\limits_{j = 1}^{n_X}\hat{\beta}_j - 1 \right|,
\end{align}
where $$\hat{\beta}_j = \frac{\text{cov}(Y_i, X_j)}{s^2_{X_j}},\ \ \ \ \ \ \ \  s^2_{X_j} \ne 0$$
and $$\hat{\alpha}_j = \bar{Y}_i - \hat{\beta}_j \bar{X}_j.$$ 

Semifast-MUOD has the advantage of reducing the computational time, since only a subsample of the functional data is used in computing the indices. Obviously, the gains in computational time is dependent on the sample size $n_X$, which is in turn dependent on the sample proportion $p$. Thus, the time complexity is reduced to an order of $\Theta(p n^2d)$. 

\subsection{Fast-MUOD}
\label{subsec:fastmuod}
For Fast-MUOD, we propose to use only the point-wise median in the computation of the indices. Let $\tilde{Y}$ be the point-wise median of the observations in $\{Y\}_{i = 1}^n$. Then, we compute the shape, magnitude and amplitude indices of any $Y_i$ w.r.t. to $\tilde{Y}$, instead of $F_Y$ or $F_X$. We define the Fast-MUOD shape index of $Y_i$ as 

\begin{equation}
I_S(Y_i, \tilde{Y}) = \left|\hat{\rho}(Y_i, \tilde{Y}) - 1 \right|.
\end{equation}
Likewise, the amplitude and magnitude indices of $Y_i$ are given by 
\begin{align}
I_A(Y_i, \tilde{Y}) &= \left|\hat\beta_i - 1 \right|\\
I_M(Y_i, \tilde{Y}) &= \left|\hat\alpha_i \right|
\end{align}
where $$\hat\beta_i = \frac{\text{cov}(Y_i, \tilde{Y} )}{s^2_{\tilde{Y}}} \ \ \ \ \ s^2_{\tilde{Y}}\ne 0,$$ and  $$\hat \alpha_i = \bar{Y_i} - \hat{\beta}_j \bar{\tilde{Y}}$$
Fast-MUOD is highly scalable since the time complexity has been reduced to an order of $\Theta(nd)$. These indices are more robust to outliers since they are computed with respect to only the point-wise median which corresponds to the depth median of the integrated Tukey halfspace depth (\citeauthor{nagyesiam}, \citeyear{nagyesiam}; \citeauthor{hubertmfhd},  \citeyear{hubertmfhd}).

\subsection{Alternative Medians and Correlation Coefficients}

The point-wise median, in general, is not necessarily one of the observed curves, and its use (in Fast-MUOD) is to create a reference ``typical" observation used for computing the indices, rather than identify a median observation of the functional data. Other median observations can be identified (and used in the computation of Fast-MUOD indices) using a functional depth measure. Although such a depth measure can also be used in detecting outliers (e.g., using a functional boxplot), our methods still provide the advantage of classifying the outliers. The point-wise median is desirable because it is fast and easy to compute, even for dense functional data. 

As an alternative, the multivariate $L_1$ median can be used. However, we have found that this is difficult to compute for dense functional data observed on lots of domain points. Moreover, the use of the $L_1$ median in computing the indices does not show any significant gains in outliers detection performance in our simulation tests, despite being more computationally expensive (see Section 1 of the Online Resource for comparison between Fast-MUOD using the $L_1$ median and the point-wise median). In general, we recommend the use of the point-wise median for dense and big functional data. For an overview of the computation of the $L_1$ median, we refer the reader to \cite{Fritz2012}.

Likewise, other robust or non-parametric correlation coefficients like Kendall's Tau and Spearman's rank correlation coefficients have been considered in the formulation of the shape indices $I_S$. Results of our tests show that the Pearson correlation coefficient provides the best outlier detection performance. See Section 4 of the Online Resource for a comparison of the performance of $I_S$ for Fast-MUOD computed using different correlation coefficients.

\subsection{Fast-MUOD and Semifast-MUOD Indices Cutoff}

After obtaining the indices (using Fast-MUOD, or Semifast-MUOD), the next step in outlier detection is to determine a cutoff value for separating the outliers from the typical observations. The theoretical distributions of these indices are unknown, but simulations show that the distributions of these indices are right skewed and that the indices of the outliers appear on the right tail. Hence, a good cutoff method should be able to find a reasonable threshold in the right tails. We propose to use a classical boxplot on the indices. We declare $Y_i$ a shape outlier if $I_S(Y_i, F) \ge Q_{3I_S} + 1.5 \times IQR_{I_S}$ where $Q_{3I_S}$  and $IQR_{I_S}$ are the third quartile and the inter-quartile range of $I_S$ respectively, for $F\in \{F_X, \tilde{Y}\}$. We apply the same cutoff rule on the magnitude and amplitude indices $I_M(Y_i, F)$ and $I_A(Y_i, F)$. The identified outliers of each type are then returned (together with their type(s)), to give a clue why they are flagged as outliers.

We have also considered other cutting methods including the transformation of the indices and the use of more specialized boxplots (e.g., the adjusted boxplot for skewed distributions of \cite{hubert2008adjusted} and the  boxplot of \cite{Carling2000}). We find that the adjusted boxplot is not sensitive enough to detect outliers and transformations of the indices usually worsen the separation between the indices of the outliers and typical observations. In our tests, the cutoff based on the classical boxplot performed well consistently acrosss the different types of outliers. Consequently, the results of the subsequent simulations and applications in this paper are obtained using this cutoff method for Semifast-MUOD and Fast-MUOD.

\begin{figure}[htbp!]
	\removelatexerror
	\begin{algorithm}[H]
		\SetAlgoLined
		$M_X = \text{sample(}M_Y, p)$ : $\mathbb{R}^{d\times n} \rightarrow \mathbb{R}^{d\times n_X}$\\
		means = colmean($M_Y$): $\mathbb{R}^{d\times n} \rightarrow \mathbb{R}^{n}$ \\
		sds = colsd($M_Y$): $\mathbb{R}^{d\times n} \rightarrow \mathbb{R}^{n}$ \\
		refmean = colmean($M_X$): $\mathbb{R}^{d\times n_X} \rightarrow \mathbb{R}^{n_X}$ \\
		refvar = colvar($M_X$): $\mathbb{R}^{d\times n_X} \rightarrow \mathbb{R}^{n_X}$ \\
		refsds = colsd($M_X$): $\mathbb{R}^{d\times n_X} \rightarrow \mathbb{R}^{n_X}$ \\
		cov = covariance($M_X$, $M_Y$): $\mathbb{R}^{d\times n_X} \times \mathbb{R}^{d\times n} \rightarrow \mathbb{R}^{n_X\times n}$\\
		cor = cov/refsds/sds : $\mathbb{R}^{n_X\times n} \times \mathbb{R}^{n_X}\times \mathbb{R}^{n}  \rightarrow \mathbb{R}^{n_X\times n}$\\
		$I_S(Y, F_X) = |\text{colmean(cor)} - 1|$ : $\mathbb{R}^{n_X\times n}  \rightarrow \mathbb{R}^{n}$\\
		$\beta$ = cov/refvar : $\mathbb{R}^{n_X\times n} \times \mathbb{R}^{n_X} \rightarrow \mathbb{R}^{n_X\times n}$\\
		$I_A(Y, F_X) = |\text{colmean(}\beta) - 1|$ : $\mathbb{R}^{n_X\times n}  \rightarrow \mathbb{R}^{n}$\\
		$\beta x$ = $\beta\times\text{refmean}$: $\mathbb{R}^{n_X\times n} \times \mathbb{R}^{n_X} \rightarrow \mathbb{R}^{n_X\times n}$\\
		
		$\alpha = \text{means} - \beta x$   : $ \mathbb{R}^{n_X\times n} \times \mathbb{R}^{n} \rightarrow \mathbb{R}^{n_X\times n}$\\
		$I_M(Y,F_X) = |\text{colmean(}\alpha) |$ : $\mathbb{R}^{n_X\times n}  \rightarrow \mathbb{R}^{n}$\\
		Return $I_A(Y, F_X), I_M(Y, F_X), I_S(Y, F_X)$
		\caption{SemiFastMUOD($M_Y$)}\label{alg:semifastmuod}
	\end{algorithm}
\end{figure}
\subsection{Implementation}
MUOD was implemented in R, (\citeauthor{Rcore} \citeyear{Rcore}) with some of the computational intensive parts of the algorithm written in $C{++}$ using the Rcpp package (\citeauthor{Rcpp} \citeyear{Rcpp}). Fast-MUOD and Semifast-MUOD follow the same implementation. 
We provide an overview into the implementation of both methods in this section. For Semifast-MUOD, $I_A(Y, F_X)$, $I_S(Y, F_X)$, and $I_M(Y, F_X)$, are computed using Algorithm \ref{alg:semifastmuod}. The algorithm takes as input the row matrix $M_Y = [Y_1, \ldots Y_n]$ built from the observations in $\{Y_i\}_{i = 1}^n$, with $|Y_i| = d$. Next, we randomly sample from the columns of $M_Y$ to create the sample row matrix $M_X = [X_1,\ldots,X_{n_X}]$, the random sample to use for computing the indices. The rest of the computation follows as outlined in  Algorithm \ref{alg:semifastmuod}. It is noteworthy that the covariance matrix in Line 7 of Algorithm \ref{alg:semifastmuod} can become quite large easily. To manage memory, we implemented the computation of the values of this matrix and the rest of the indices sequentially in $C{++}$, so that we do not have to store the covariance matrix in memory. The implementation for Fast-MUOD is very similar and is outlined in Algorithm \ref{alg:fastmuod}. The algorithm takes as input $M_Y$ and then computes the point-wise median $\tilde{Y} \in \mathbb{R}^{d}$ which is used in the computation of the indices. The operations ``colmean($\cdot$)", ``colmedian($\cdot$)", ``colsd($\cdot$)", and ``colvar($\cdot$)" used in both algorithms indicate column-wise mean, median, standard deviation, and variance operations respectively.

\begin{figure}[htbp!]
	\removelatexerror
	\begin{algorithm}[H]
		\SetAlgoLined
		$\tilde{Y} = \text{colmedian}(M_Y)$: $\mathbb{R}^{d\times n} \rightarrow \mathbb{R}^{d}$ \\
		means = colmean($M_Y$): $\mathbb{R}^{d\times n} \rightarrow \mathbb{R}^{n}$ \\
		sds = colsd($M_Y$): $\mathbb{R}^{d\times n} \rightarrow \mathbb{R}^{n}$ \\
		refmean = mean($\tilde{Y}$): $\mathbb{R}^{d} \rightarrow \mathbb{R}$ \\
		refvar = var($\tilde{Y}$): $\mathbb{R}^{d} \rightarrow \mathbb{R}$ \\
		refsds = sd($\tilde{Y}$): $\mathbb{R}^{d} \rightarrow \mathbb{R}$ \\
		cov = covariance($\tilde{Y}$, $M_Y$): $\mathbb{R}^{d} \times \mathbb{R}^{d\times n} \rightarrow \mathbb{R}^{n}$\\
		cor = cov/refsds/sds : $\mathbb{R}^{n} \times \mathbb{R}\times \mathbb{R}^{n} \times \rightarrow \mathbb{R}^{n}$\\
		$I_S(Y, \tilde{Y}) = |\text{cor} - 1|$ : $\mathbb{R}^{n}  \rightarrow \mathbb{R}^{n}$\\
		$\beta$ = cov/refvar : $\mathbb{R}^{n} \times \mathbb{R}  \rightarrow \mathbb{R}^{n}$\\
		$I_A(Y, \tilde{Y}) = |\beta - 1|$ : $\mathbb{R}^{n}  \rightarrow \mathbb{R}^{n}$\\
		$\beta x$ = $\beta\times\text{refmean}$: $\mathbb{R}^{n} \times \mathbb{R} \rightarrow \mathbb{R}^{n}$\\
		
		$\alpha =  \text{means} - \beta \cdot x $   : $ \mathbb{R}^{ n} \times \mathbb{R}^{n} \rightarrow \mathbb{R}^{n}$\\
		$I_M(Y, \tilde{Y}) = |\alpha|$ : $\mathbb{R}^{n}  \rightarrow \mathbb{R}^{n}$\\
		Return $I_A(Y, \tilde{Y}), I_M(Y, \tilde{Y}), I_S(Y, \tilde{Y})$
		\caption{FastMUOD($M_Y$)}\label{alg:fastmuod}
	\end{algorithm}
\end{figure}

\section{Simulation Study}
\label{sec:simstudy}
In this section, we evaluate the performance of the proposed methods using some simulation experiments. 

\subsection{Outlier Models}
\label{subsec::outmodel}
In our simulation study, we generate curves from different outlier models that have been studied in \cite{dai2018multivariate}, \cite{arribas2014shape}, \cite{febrero2008} and \cite{sun2011functional}. In total, we consider eight models where the first model, Model 1, is a clean model with no outlier, while Models 2 -- 8 contain outliers. The base models and the corresponding contamination models are specified below.

\begin{figure*}[htbp!]
	\centering
	\includegraphics[scale = .70]{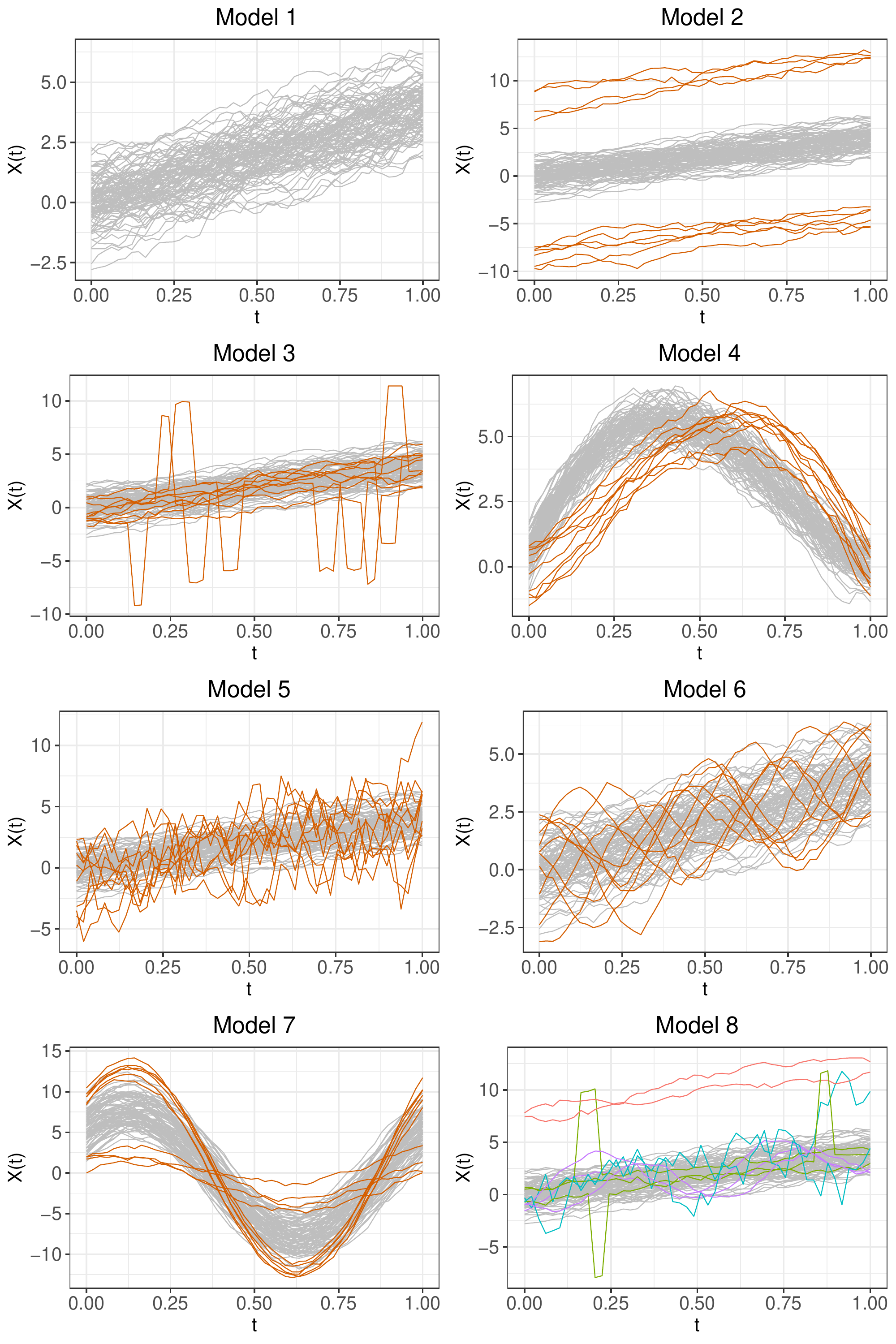}
	\caption{Sample data generated by the eight simulation models ($\alpha= 0.10$, $n = 100$ and $d = 50$). Outliers are in color.}
	\label{fig:model_illus}
\end{figure*}

\begin{itemize}
	\item \textbf{Model 1}: Main model $X_i(t) = 4t + e_i(t)$ with no contamination, for $i = 1, \ldots,n$. $e_i(t)$ is a Gaussian process with zero mean and covariance function $\gamma(s,t) = \exp\{-|t-s|\}$, where $s,\ t \in [0, 1]$.
	
	\item \textbf{Model 2}: Main model: same as Model 1; Contamination model: $X_i(t) = 4t + 8k_i + e_i(t)$, for $i  = 1, \ldots, n$ and $k_i \in \{-1, 1\}$ with equal probability. $e_i(t)$ remains as defined above. This is a shifted model where the generated curves are magnitude outliers shifted from the main model.
	
	\item \textbf{Model 3}: Main model: same as Model 1; Contamination model: $X_i(t) = 4t + 8k_iI_{T_i \le t\le T_i+0.05 } + e_i(t)$, for $i  = 1, \ldots, n$, $T_i \sim \text{Unif}(0.1, 0.9)$, and $I$ an indicator function. $k_i$ and $e_i(t)$ remain as defined above. The outlying curves from this model are magnitude outliers for only a small portion of the domain, which produce spikes along the domain.
	
	\item \textbf{Model 4}: Main model:  $X_i(t) = 30t(1-t)^{3/2} + \bar{e_i}(t)$; Contamination model: $X_i(t) = 30t^{3/2}(1-t) + \bar{e_i}(t)$, for $i  = 1, \ldots, n$; where $\bar{e_i}(t)$ is a Gaussian process with zero mean and covariance function $\bar{\gamma}(s,t) = 0.3\exp\{-|s-t|/0.3\}$ with $s, t \in [0,1]$. The outlying curves in this model produces outliers that are similar to typical observations but slightly shifted horizontally and reversed.

	\item \textbf{Model 5}: Main model: same as Model 1; Contamination model: $X_i(t) = 4t + e_{2_i}(t)$, for $i  = 1, \ldots, n$; where $e_{2_i}(t)$ is a Gaussian process with zero mean and covariance function $\gamma_2(s,t) = 5\exp\{-2|t-s|^{0.5}\}$ with $s, t \in [0,1]$. The outlying curves generated are shape outliers with a different covariance function even though they follow the general 
	trend of the normal observations.
	
	\item \textbf{Model 6}: Main model: Same as Model 1, Contamination model: $X_i(t) = 4t +2\sin(4(t + \theta_i)\pi) + e_i(t)$, for $i  = 1, \ldots, n$; where $\theta_i \sim Unif(.25, .75)$. $e_i(t)$ remains as defined above. Like Model 5 above, the generated outlying curves have the same trend as the normal observations but they have a different covariance function that is periodic in nature. 
	
	\item \textbf{Model 7}: Main model  $X_i(t) = a_i \sin \theta + b_i \cos \theta + e_i(t)$; Contamination model: $X_i(t) = (9 \sin \theta + 9\cos \theta)\cdot (1 - u_i) + (p_i \sin \theta + q_i \cos \theta) u_i + e_i(t)$, for $i  = 1, \ldots, n$; where $\theta \in [0, 2\pi]$, $a_i, b_i \sim \text{Unif}(3, 8)$, $p_i, q_i \sim $ $\text{Unif}(1.5, 25.)$ and $u_i\in \{0, 1\}$ with equal probability. $e_i(t)$ remains as defined above. The contaminating curves are amplitude outliers with a similar periodic shape as the normal observations but with slightly increased or decreased amplitude. 
	
	\item \textbf{Model 8}:  Main model: same as Model 1;
	Contamination model: For each outlier to be generated, a contamination model is sampled from any of the following contamination models (with equal probability):
	\begin{enumerate}
		\item Contamination model of Model 2
		\item Contamination model of Model 3
		\item Contamination model of Model 5
		\item Contamination model of Model 6
	\end{enumerate}
	Thus, Model 8 is a mixture model containing different types of outliers. 
	
\end{itemize}
Simulated data from these eight models will be a mixture of observations from the main model with outliers from the contamination model, where number of outliers is determined by the contamination rate alpha $\alpha$.  In the subsequent simulation results, we set the contamination rate $\alpha = 0.1$ for each Model 2 -- 8, and we generate $n = 300$ curves on $d = 50$ equidistant points on the interval $[0,1]$.  Figure \ref{fig:model_illus} shows a sample of the eight models with $\alpha= 0.1$, $n = 100$ and $d = 50$. 

\subsection{Outlier Detection Methods}
\label{subsec::competing-methods}
We focus on comparing the outlier detection performance of the proposed methods to MUOD and to other recent outlier detection methods for functional data. Since the proposed methods produce three types of outliers (magnitude, amplitude, and shape), there are different ways to study the performance of the methods. For instance, a user might decide to target only magnitude outliers and discard the other types of outliers (produced by the methods) based on practical background and use case scenario. On the other hand, one might decide to consider all the outliers provided by the proposed methods. Consequently, we will consider the following different possible sets in our comparison:

\begin{itemize}
	\item FST: This is the union of the different types of outliers flagged by Fast-MUOD. Thus, an observation is an outlier under this scheme if it is either a shape, magnitude or amplitude outlier. In our simulation, we used the pointwise median for Fast-MUOD but the results obtained with the $L_1$ median are similar (see Section 1 of the Online Resource). 
	\item FSTMG: This considers only the magnitude outliers flagged by Fast-MUOD. Consequently, an observation is an outlier only if it is flagged by Fast-MUOD as a magnitude outlier. 
	\item FSTSH: This considers only the shape outliers flagged by Fast-MUOD. Thus, an observation is an outlier only if it is flagged by Fast-MUOD as a shape outlier.
	\item FSTAM: This is considers only the amplitude outliers flagged by Fast-MUOD. An observation is an outlier only if it is flagged by Fast-MUOD as an amplitude outlier.
	\item SF: This is the union of the different types of outliers flagged by Semifast-MUOD (using a random sample whose size is 50\% of the size of the original data).
	\item SF25: This is the union of the different types of outliers flagged by Semifast-MUOD but using a random sample whose size is 25\% of the size of the original data.  
	\item MUOD: This is the union of the different types of outliers flagged by MUOD as proposed in \cite{azcorra2018unsupervised} (using the ``\textit{tangent method}" to determine a cutoff).  
\end{itemize}

Considering the different types of outliers flagged by Fast-MUOD  in our performance evaluation gives a clear picture of how the different types of outliers contribute to the overall performance of Fast-MUOD (FST). It is easy to do the same for SF and SF25. However, we do not include these results here but rather their overall performance since the results are quite similar to those of Fast-MUOD. We compare the methods above to the following outlier detection methods for functional data.

\begin{itemize}
	\item OGMBD: The outliergram method, proposed in \cite{arribas2014shape}, mainly targets shape outliers. It uses a scatter plot of the modified band depth (MBD) and the modified epigraph index (MEI). Outliers are identified by using a boxplot to find the most distant points that lie below the parabola generated by the plot of (MEI, MBD). In addition, outliergram uses the functional boxplot to detect magnitude outliers. Thus, in our evaluation, we consider outliers flagged by both outliergram and functional boxplot.   
	
	\item MSPLT: MS-Plot is based on a directional outlyingness for multivariate functional data proposed by \cite{dai2019directional} and it decomposes the ``total directional outlyingness" of sample curves into ``magnitude outlyingness" (represented by the ``mean directional outlyingness", MO) and ``shape outlyingness" (represented by the ``variation of directional outlyingness", VO). The MS-Plot is then the scatter plot of $(\text{MO}, \text{VO})^\text{T}$. Outlying curves are identified by computing the squared robust Mahalanobis distance of $(\text{MO}, \text{VO})^\text{T}$ (using the minimum covariance determinant (MCD) algorithm of \cite{rousseeuw1999fast}), and approximating the distribution of these distances using an $F$ distribution according to \cite{hardinrocke3005}. Curves with robust distance greater than a threshold obtained from the tails of the $F$ distribution are flagged as outliers.    
	
	\item TVD: This method uses the total variation depth (TVD) proposed by \cite{huang2019decomposition} to compute a ``(modified) shape similarity" (MSS) index of the sample functions. A classical boxplot, with the $F_c \times IQR$ cutoff rule (where $F_c$ is the factor), is then applied on the MSS index to detect shape outliers. After removing the shape outliers, a functional boxplot (using TVD to construct the central region) is then applied on the remaining observation in order to detect magnitude outliers. The magnitude outliers are functions outside of 1.5 times the 50\% central region with respect to the original sample before the shape outliers were removed. In our simulation study, we used the default value of $F_c = 3$.
	\item FOM: The functional outlier map, proposed by \cite{rousseeuw2018measure},  uses a ``directional outlyingness" (DO) measure. This measure is then extended to functional data to get the ``functional directional outlyingnees" (fDO), computed at the observed points of each function's domain. The variability of the DO values (vDO), is then defined, and the FOM is the scatterplot of (fDO, vDO). To flag observations as outliers, the ``combined functional outlyingness" (CFO), based on fDO and vDO, is computed, transformed to logarithm (LCFO), and standardized in a robust way (SLCFO). Any observation with $SLCFO > \Phi^{-1}(.995)$ is then flagged as an outlier, where $\Phi(\cdot)$ is the standard normal cumulative distribution.
	
	\item FAO: The functional adjusted outlyingness is similar to FOM above, but uses the ``adjusted outlyingness" (AO) proposed by \cite{brys2005robustification} (see also \cite{hubert2008outlier} and \cite{hubert2015multivariate}) instead of the DO proposed in  \cite{rousseeuw2018measure}. The AO can be extended to functional data to get a functional Adjusted Outlyingness (fAO). The variability (vAO) of the fAO and fAO itself can then be used in a scatterplot to build a functional outlier map as done in FOM above. 

\item FOM2 and FAO2: Since the functional directional outlyingness and functional adjusted outlyingness can be computed for multivariate functional data, we add the first derivatives of the simulated data as a second dimension to the original data and analyse the obtained bivariate functional data with functional outlier maps of fDO (for FOM2) and fAO (for FAO2).	
	\item ED: The extremal depth notion proposed by \cite{narisetty2016extremal} orders functions using a left-tail stochastic ordering of the depth distribution. This depth notion focuses mainly on ``extreme outlingness" just as the name implies, and thus tends to penalize functions with extreme values, even if these extreme values occur in small portions of the domain. In our simulation, we use the extremal depth to construct a central region which is used in a functional boxplot to detect outliers.
	\item SEQ1, SEQ2 and SEQ3: These methods detect outliers using some standard sets of sequential transformations proposed in \cite{dai2020sequential}. The functional data are sequentially transformed and outliers are removed after each transformation using a functional boxplot based on some depth measure. The first transformation proposed is $\{\mathcal{T}_0(Y_i)\}_{i = 1}^n$ and it indicates applying a functional boxplot to the raw data $\{Y_i\}_{i = 1}^n$ (to get the $\mathcal{T}_0-$outliers). Other proposed transformation include shifting the curves $\{Y_i\}_{i = 1}^n$ to their centers: 
	$$\mathcal{T}_1(Y_i(t)) = Y_i(t) - \lambda (\mathcal{I})^{-1}\int_{\mathcal{I}}Y_i(t)dt,$$
	where $\lambda (\mathcal{I})$ is the Lebesgue measure of $\mathcal{I}$; and normalising the curves $\{Y_i\}_{i = 1}^n$ with their $L_2$ norm:
	$$\mathcal{T}_2(Y_i(t)) = Y_i(t)\lvert\lvert Y_i(t) \rvert\rvert_2^{-1},$$ with $\lvert\lvert Y_i(t) \rvert\rvert_2 = \left[\int_{\mathcal{I}}\{Y_i(t)\}^2dt\right]^{1/2}$. Additional transformations involve taking the first order derivatives of the raw curves (denoted by $\mathcal{D}_1$) and further differentiating the first order derivatives (denoted by $\mathcal{D}_2$). 
	
	We consider the following sequence of transformations: 
	\begin{enumerate}
	\item SEQ1 $= \{\mathcal{D}_1\circ\mathcal{T}_1\circ \mathcal{T}_0(Y_i)\}_{i = 1}^n$,
	\item SEQ2 $= \{\mathcal{T}_2\circ\mathcal{T}_1\circ \mathcal{T}_0(Y_i)\}_{i = 1}^n$,
	\item SEQ3 $= \{\mathcal{D}_2\circ\mathcal{D}_1\circ \mathcal{T}_0(Y_i)\}_{i = 1}^n$.
	\end{enumerate}
	We use the distance based $L^{\infty}$ functional depth \citep{long2015study} for ordering the curves in the intermediate functional boxplots applied after each transformation. We selected $L^{\infty}$ depth because it had one of the best performance in the simulation study conducted in \cite{dai2020sequential}. 
	
\end{itemize}

\begin{table*}
\ra{.95}
\centering
\caption{\label{tab:sim-results-300-50}Mean and Standard Deviation (in parentheses) of the True Positive Rates (TPR) and False Positive Rate (FPR) over eight simulation models with 500 repetitions for each possible case. Each simulation is done with $n=300$ and $d=50$ and $\alpha = 0.1$. Comparatively high TPRs are in bold. Proposed methods in italics.}
{\scriptsize	
\begin{tabular}{@{}lcccccccc@{}}  \toprule
  			\multirow{2}{*}{Method} & \multicolumn{2}{c}{Model 1}  & \multicolumn{2}{c}{Model 2} & \multicolumn{2}{c}{Model 3} & \multicolumn{2}{c}{Model 4} \\
  			\cmidrule{2-3} \cmidrule{4-5} \cmidrule{6-7} \cmidrule{8-9}
  			& TPR & FPR & TPR & FPR & TPR & FPR & TPR & FPR \\ 
  			\midrule
\textit{FST}   & -& 9.90(1.50) & \textbf{100.00(0.00)} & 8.95(1.59) & \textbf{99.81(0.89)} & 6.10(1.37) & \textbf{100.00(0.00)} & 3.15(1.13) \\ 
\textit{FSTMG} & -& 1.74(0.92) & \textbf{99.99(0.15)} & 0.36(0.40) & 4.13(3.70) & 1.52(0.89) & 41.25(9.71) & 0.63(0.54) \\ 
\textit{FSTSH} & -& 7.94(1.34) & 7.79(4.86) & 7.94(1.48) & \textbf{98.97(2.03)} & 4.36(1.12) & \textbf{100.00(0.00)} & 2.24(0.95) \\ 
\textit{FSTAM} & -& 1.70(0.83) & 1.66(2.43) & 1.69(0.93) & 6.41(4.62) & 1.38(0.80) & 54.56(11.76) & 0.45(0.45) \\ 
\textit{SF}    & - & 9.58(1.54) & \textbf{100.00(0.00)} & 8.66(1.53) & \textbf{99.49(1.49)} & 5.60(1.32) & \textbf{99.94(0.44)} & 2.65(1.02) \\ 
\textit{SF25}  & - & 9.60(1.51) & \textbf{99.99(0.21)} & 8.63(1.58) & \textbf{99.53(1.29)} & 5.59(1.28) & \textbf{99.87(0.67)} & 2.59(1.03) \\ 
\textit{MUOD} & - & 12.07(4.41) & \textbf{99.75(3.26)} & 8.40(3.78) & 56.32(24.44) & 10.67(4.95) & 95.39(10.02) & 3.95(2.83) \\ 
 OGMBD  & -& 4.77(1.25) & \textbf{100.00(0.00)} & 4.65(1.35) & 39.43(11.40) & 3.45(1.16) & 93.54(5.21) & 1.22(0.72) \\ 
MSPLT & -& 3.72(1.41) & \textbf{99.97(0.33)} & 2.90(1.24) & \textbf{100.00(0.00)} & 2.95(1.34) & \textbf{99.95(0.39)} & 1.36(0.84) \\ 
TVD & -& 0.00(0.03) & \textbf{100.00(0.00)} & 0.00(0.03) & \textbf{100.00(0.00)} & 0.00(0.00) & 2.77(3.77) & 0.00(0.00) \\ 
FOM & - & 0.53(0.52) & \textbf{100.00(0.00)} & 0.07(0.17) & 47.11(18.59) & 0.09(0.19) & 48.23(17.44) & 0.06(0.15) \\ 
FAO & - & 0.21(0.34) & \textbf{100.00(0.00)} & 0.02(0.08) & 25.97(16.11) & 0.02(0.09) & 5.92(8.13) & 0.02(0.09) \\ 
FOM2 & - & 3.96(1.13) & \textbf{100.00(0.00)} & 1.26(0.73) & \textbf{100.00(0.00)} & 2.02(0.95) & 78.98(8.93) & 0.78(0.58) \\ 
FAO2 & - & 3.39(1.17) & \textbf{100.00(0.00)} & 1.41(0.83) & \textbf{100.00(0.00)} & 1.59(0.89) & 32.31(15.17) & 0.80(0.62) \\ 
ED & - & 0.00(0.00) & \textbf{99.99(0.15)} & 0.00(0.02) & \textbf{99.09(1.73)} & 0.00(0.00) & 0.12(0.62) & 0.00(0.00) \\ 
SEQ1 & - & 0.00(0.01) & \textbf{99.99(0.15)} & 0.00(0.00) & \textbf{100.00(0.00)} & 0.00(0.00) & 10.49(7.41) & 0.00(0.00) \\ 
SEQ2 & - & 0.65(0.46) & \textbf{99.99(0.15)} & 0.68(0.49) & \textbf{100.00(0.00)} & 0.61(0.48) & 29.56(13.57) & 0.00(0.00) \\ 
SEQ3 & - & 0.00(0.00) & \textbf{99.99(0.15)} & 0.00(0.00) & \textbf{100.00(0.00)} & 0.00(0.00) & 5.58(4.72) & 0.00(0.00) \\ 
\bottomrule
\multirow{2}{*}{Method} & \multicolumn{2}{c}{Model 5} & \multicolumn{2}{c}{Model 6} & \multicolumn{2}{c}{Model 7} & \multicolumn{2}{c}{Model 8} \\
			\cmidrule{2-3} \cmidrule{4-5} \cmidrule{6-7} \cmidrule{8-9}
			& TPR & FPR & TPR & FPR & TPR & FPR & TPR & FPR \\ 
			\midrule
\textit{FST} & \textbf{95.97(4.27)} & 5.67(1.19) & \textbf{93.05(6.42)} & 6.31(1.35) & \textbf{79.73(14.95)} & 6.55(1.91) & \textbf{98.63(2.45)} & 6.65(1.40) \\ 
\textit{FSTMG} & 15.94(6.70) & 1.08(0.71) & 0.83(1.70) & 1.77(0.94) & 1.65(2.36) & 1.69(0.90) & 30.65(8.10) & 1.04(0.75) \\ 
\textit{FSTSH} &  86.35(6.74) & 4.39(1.12) & \textbf{91.01(6.75)} & 4.35(1.10) & 4.21(3.68) & 4.94(1.72) & 71.77(7.58) & 5.31(1.29) \\ 
\textit{FSTAM} &  22.99(7.93) & 1.01(0.71) & 3.54(3.78) & 1.40(0.79) & \textbf{79.10(15.42)} & 0.01(0.05) & 10.74(5.69) & 1.29(0.83) \\ 
\textit{SF} &  \textbf{94.05(5.00)} & 5.27(1.22) & \textbf{92.46(6.31)} & 5.87(1.32) & 67.31(17.06) & 6.54(1.86) & \textbf{98.11(2.69)} & 6.19(1.35) \\ 
\textit{SF25}&  \textbf{93.70(5.44)} & 5.26(1.25) & \textbf{91.95(6.74)} & 5.84(1.25) & 66.75(17.67) & 6.63(1.92) & \textbf{97.85(2.97)} & 6.15(1.33) \\ 
\textit{MUOD}&  50.16(14.96) & 4.35(2.96) & 48.60(23.31) & 12.01(5.45) & \textbf{98.22(5.95)} & 18.77(6.15) & 64.23(16.18) & 4.41(3.25) \\ 
OGMBD  & \textbf{95.99(4.02)} & 1.97(0.88) & \textbf{99.89(0.60)} & 1.88(0.89) & 15.40(14.65) & 0.00(0.00) & 82.99(7.39) & 2.87(1.16) \\ 
MSPLT &  \textbf{99.99(0.15)} & 2.81(1.21) & \textbf{100.00(0.00)} & 2.91(1.26) & 66.39(16.04) & 0.02(0.08) & \textbf{99.98(0.26)} & 2.85(1.31) \\ 
TVD &  \textbf{100.00(0.00)} & 0.00(0.02) & 84.25(12.13) & 0.00(0.02) & 40.88(12.82) & 0.00(0.02) & \textbf{99.42(1.51)} & 0.00(0.00) \\ 
FOM &  10.84(7.76) & 0.09(0.21) & 0.02(0.26) & 0.08(0.18) & 0.64(1.72) & 0.00(0.03) & 40.85(10.30) & 0.10(0.21) \\ 
FAO&  7.11(6.55) & 0.03(0.10) & 0.02(0.26) & 0.02(0.08) & 0.57(1.67) & 0.00(0.03) & 33.20(9.53) & 0.03(0.12) \\ 
FOM2  & \textbf{100.00(0.00)} & 2.06(0.94) & 71.25(15.41) & 1.90(0.86) & 30.75(16.49) & 0.17(0.25) & \textbf{99.01(2.05)} & 1.92(0.89) \\ 
FAO2 &  \textbf{100.00(0.00)} & 1.61(0.83) & 54.90(17.75) & 1.47(0.83) & 12.88(11.47) & 0.11(0.22) & \textbf{97.06(3.47)} & 1.56(0.84) \\ 
ED &  25.01(9.20) & 0.00(0.00) & 0.01(0.15) & 0.00(0.00) & 0.00(0.00) & 0.00(0.00) & 55.85(8.85) & 0.00(0.00) \\ 
SEQ1  & \textbf{100.00(0.00)} & 0.00(0.00) & 0.17(0.76) & 0.00(0.00) & 0.00(0.00) & 0.00(0.00) & 75.55(7.86) & 0.00(0.02) \\ 
  SEQ2  & 83.91(7.41) & 0.57(0.46) & 6.90(5.50) & 0.60(0.46) & 1.61(2.26) & 0.00(0.03) & 74.68(7.80) & 0.60(0.45) \\ 
  SEQ3  & \textbf{100.00(0.00)} & 0.00(0.00) & 0.00(0.00) & 0.00(0.00) & 0.00(0.00) & 0.00(0.00) & 74.87(7.87) & 0.00(0.02) \\ 
   \bottomrule
\end{tabular}}
\end{table*}

\subsection{Simulation Results}

For each of the Models 2 -- 8, we evaluate the true positive rate (TPR), the percentage of correctly identified out of the true outliers, and the false positive positive rate (FPR), the percentage of false positives out of the number of non-outliers. Since Model 1 is a clean model, we present only the FPR under Model 1. 

Table \ref{tab:sim-results-300-50} shows the mean and standard deviation (in parenthesis) of the TPRs and FPRs for all the methods over 500 repetitions. In Model 1, where we have a clean model, MUOD has an exceptionally high FPR of $12.07\%$ mainly because of the aggressive tangent cutoff method which it uses for detecting outliers. FST, SF, and SF25 which use the classical boxplot as a cutoff technique show better FPRs than MUOD. Compared to other functional outlier detection methods, the proposed MUOD-based methods (FST, SF and SF25) show higher FPRs. This is because FST, SF and SF25 are the unions of the three types of outliers flagged by Fast-MUOD, Semifast-MUOD, and Semifast-MUOD with 25\% of the sample, respectively. For instance, FST is the union of FSTMG, FSTSH and FSTAM, and consequently, it inherits the FPRs of these individual methods (same applies for SF and SF25). However,  considering the individual methods, FSTMG and FSTAM, we see low FPRs. In fact, the overall FPRs of the MUOD-based methods are mainly driven by the FPR of the shape outliers (as seen with FST and FSTSH). This is because the MUOD based methods use a simple Pearson correlation as an index to identify shape outliers and this might be affected to some extent by random noise. However, we find this not be too much of an issue in real life use cases, especially because it is typical for functions to be smoothed (or represented with some basis function with implicit smoothing effect) during the exploratory analysis process (for example, see Section \ref{sec::app}).

All the methods have very high accuracy for Model 2 where we have magnitude outliers, except for FSTSH and FSTAM which target shape and amplitude outliers respectively. FST, SF and SF25 have higher FPRs for the same reasons explained above for Model 1. However, FSTMG has a very low FPR of 0.36\%, a value comparable to or better than some other methods. For Models 3 - 8, which contain either shape, amplitude or a mixture of outliers, the proposed MUOD-based methods (FST, SF and SF25) show good outlier detection performance. Considering the individual outliers flagged by FST across these models, the performance of FSTMG, FSTSH and FSTAM vary depending on the type of outliers contained in the model, showing the effectiveness of these individual methods in targeting their specific types of outliers. However, FOM and FAO have very low TPRs for models with shape or amplitude outliers, indicating that they are only well suited to identifying magnitude outliers. FOM2 and FAO2 have high accuracy on Models 3 and 5 as they analyse a bivariate data which includes the first derivative of the simulated data. They however struggle with Models 4, 6 and 7 which contain shape and amplitude outliers. In Model 3, FST shows a very good TPR of 99.81\%, mostly buoyed by the outliers detected by FSTSH with its TPR of 98.97\%. Furthermore, MSPLT and TVD performed excellently on Model 3. The same applies to ED, with its emphasis on extreme outlyingness of functions, even if such outlyingness is within a small portion of the domain, a property that Model 3 clearly satisfies. OGMBD and MUOD, on the other hand, have very low TPRs for Model 3. For Model 4 however, TVD and ED fail with very low TPRs while MSPLT and FST showed excellent performance. The outliers detected by FSTSH (100.00\% TPR) in this model contributed to the overall performance of FST (the same applies for SF and SF25). OGMBD and MUOD also show very good outlier detection performance on Model 4 with 93.54\% and 95.39\% TPRs respectively. In Model 6, TVD did not have quite as good TPR compared to MSLPT, OGMB and the MUOD-based methods even though this model contains pure shape outliers. The methods based on sequential transformations (SEQ1, SEQ2 and SEQ3) have good outlier detection  performance on Models 2, 3 and 5 but they are ineffective on Models 6 and 7 which contain pure shape and amplitude outliers, respectively.

Only FST gives a satisfactory outlier detection performance for Model 7 with its TPR of 79.73\%. This model contains pure amplitude outliers and it is especially challenging because the outliers are quite similar in shape and magnitude to the non-outliers. The amplitude outliers detected by FSTAM helped FST to have this satisfactory performance. SF and SF25 on the other hand have low TPRs. Since these two methods use a random sample of the data (50\% and 25\% of sample size for SF and SF25 respectively), they are not as sensitive as FST which uses only the the point-wise (or $L_1$) median. This proves to be an advantage in detecting outliers that are very similar in shape and magnitude to non-outliers. MUOD, on the other hand has a high TPR but also a very high FPR of 18.34\% (caused by the tangent cutoff method) which makes the overall performance bad. Finally, the proposed MUOD-based methods show a good outlier detection performance on Model 8 which contains a mixture of pure shape and magnitude outliers. Likewise MSPLT and TVD show good outlier detection TPRs on Model 8, while MUOD and ED have very low TPRs. OGMB, on the other hand, did not have quite as high TPR compared to MSPLT, TVD and the MUOD-based methods. 

While we do not claim that the proposed methods are capable of identifying every possible type of outlier, the results of the simulation study has shown that the proposed MUOD-based methods (and especially Fast-MUOD) have a good and well-balanced performance over a wide range of different outlier types, thanks to the fact that they target three different types of outliers simultaneously. Since the outliers identified are also classified into different types (magnitude, amplitude or shape), additional information is provided to the user as to possible reasons why an outlier is indeed flagged as such without the need for manual inspection or data visualization. This will prove valuable when exploring large functional datasets (where visualization is difficult) and also enables selective targeting of different outlier types based on practical background and use case.

\subsection{Computational Time}
A major advantage of the proposed methods is their simplicity despite their effectiveness. Since the indices for Fast-MUOD are very easy to compute, the computational overhead of Fast-MUOD is quite low compared to other existing outlier detection methods. For functional datasets of typical size, this does not matter much. However, a fast method is required in order to handle large (and dense) functional datasets and Fast-MUOD excels in this regard (e.g., see Subsection \ref{subsec:surveillance}). While Semifast-MUOD is not as fast compared to Fast-MUOD, it is faster than the original MUOD as described in \cite{azcorra2018unsupervised}, while achieving better outlier detection performance. 

In this section, we focus on comparing the running times of the proposed methods to MUOD and to other existing outlier detection methods used in Subsection \ref{subsec::outmodel}. We generated data from Model 2 (in Subsection \ref{subsec::outmodel}) with the number of observations $n \in \{10^2, 3\times 10^2, 10^3, 3\times 10^3, 10^4, 3\times 10^4, 10^5\}$. To generate the data, we set $d = 100$ and contamination rate $\alpha = 0.05$. We used the $tictoc$ package in R (\citeauthor{tictoc} \citeyear{tictoc}) to get the running time of each method. For each $n$, we ran $20$ iterations and took the median. The experiment was run on a computer with a Core i9 8950HK processor (6 cores, 12 threads, up to 4.8GHz) with 32GB RAM.

\begin{figure*}[htbp!]
	\centering
	\includegraphics[width=.9\linewidth]{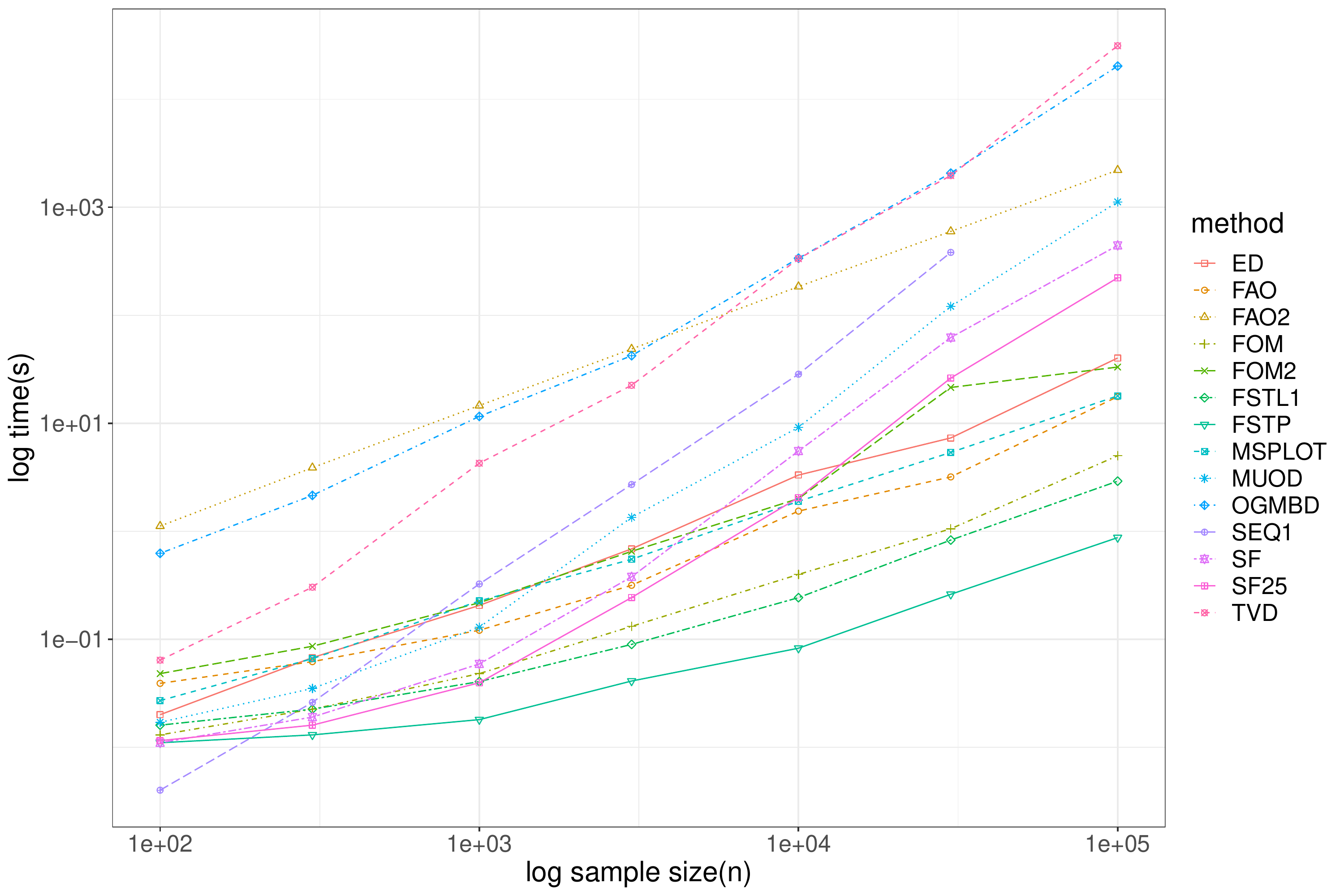}
	\caption{Plot of the median computational time of the different outlier detection methods in log-log axes. Each simulation  is  done  with $d=  100$ and $\alpha = 0.05$ with data generated from Model 2. Legend: \textit{FSTP}: Fast-MUOD computed with point-wise median, \textit{FSTL1}: Fast-MUOD computed with the $L_1$ median.}
	\label{fig:comp_time}
\end{figure*}

Figure \ref{fig:comp_time} shows the results of the running time of the different methods with log-log axes. For small sample size, all the methods have relatively short running time. However, for larger sample sizes, FSTP and FSTL1 (representing Fast-MUOD with the point-wise median and $L_1$ median respectively) have the shortest running time taking just about 0.8 and 2.9 seconds respectively to process 100,000 observations. ED, FOM, FAO and MSPLOT also show reasonable running times for large number of observations. TVD and OGMBD however are quite slow, requiring over 5 and 8 hours respectively to handle 100,000 observations. Consequently, these methods are only suitable for relatively small data (see Subsection \ref{subsec:surveillance} for example). SF and SF25 are much faster than MUOD taking about 7 and 3 minutes respectively for 100,000 observations (compared to the 18 minutes required by MUOD). The methods based on sequential transformation have similar running times and hence we show only the results for SEQ1 in Figure \ref{fig:comp_time}. These methods take about 6 minutes to process 30,000 observations (we could not run the tests up to 100,000 observations on these methods due to memory issues). Finally, FAO2 and FOM2 take about 36 minutes and 33 seconds respectively to handle 100,000 observations.

An alternative way to evaluate computational time is to evaluate the maximum number of observations a method can handle within a given set time. Using the same setup as before (simulated data from Model 2, with $d = 100$ and contamination rate $\alpha = 0.05$), we evaluate the maximum number of observation that each method can handle under 10 seconds. Starting from a sample size of 100 to 10,000, we increase the number of observations in steps of 100, while from 20,000 up to 2 million, we increase the sample size in steps of 10,000. Table \ref{tab:comp-time} below shows the result. FSTP handles over 1 million observations in less than 10 seconds while FSTL1 handles 290,000 observations under 10 seconds. Given the significant difference in computational time between FSTP and FSTL1 (due to the computation of the $L_1$ median), we recommend to always use FSTP for large data since the outlier detection performance for both methods are quite similar as mentioned earlier. Compared to other methods, FOM can handle only 270,000 observations, while MSPLOT can handle only 50,000 observations under 10 seconds. As expected, OGMBD and TVD, with their slow running times, can only handle about 1,000 and 2,000 observations respectively under 10 seconds. 

The codes for OGMBD and MSPLOT used in this experiment were obtained from the supplementary materials of \cite{arribas2014shape} and \cite{dai2018multivariate} respectively, while the implementation of TVD used is at \href{https://github.com/hhuang90/TVD}{github.com/hhuang90/TVD} as stated in \cite{huang2019decomposition}. FOM, FAO, FOM2 and FAO2 were based on codes obtained from \href{http://wis.kuleuven.be/stat/robust/software}{wis.kuleuven.be/stat/robust/software} while code for ED was obtained from the authors of \cite{narisetty2016extremal}. SEQ1, SEQ2 and SEQ3 are based on their respective implementations in the \textit{fdaoutlier} R package \citep{fdaoutlier}. Worthy of note is that the implementations of these methods used in this experiment might not be the most optimal version available (or attainable) of the respective methods.

\begin{table}[htbp!]
	\centering
	\caption{Number of observations handled under 10 seconds. Simulated data from Model 2, with $d = 100$ and contamination rate $\alpha = 0.05$}
	\label{tab:comp-time}
	{\renewcommand{\arraystretch}{0.9}
		\footnotesize
		\begin{tabular}{lrc}
			\hline
			Method & Sample size & Time (s) \\ 
			\hline
			FSTP & 1,060,000 & 9.89 \\
			FSTL1 & 290,000 & 9.41 \\ 
			SF & 10,000 & 7.36 \\ 
			SF25 & 20,000 & 9.25 \\ 
			MUOD & 10,000 & 9.58 \\ 
			OGMBD & 1,000 & 8.06 \\ 
			MSPLOT & 50,000 & 9.37 \\ 
			TVD & 2,000 & 9.97 \\ 
			FOM & 270,000 & 9.99 \\ 
			FAO & 90,000 & 9.86 \\ 
			FOM2 & 40000 & 9.32 \\ 
			FAO2 & 700 & 9.86 \\ 
			ED & 40,000 & 9.45 \\ 
			SEQ1 & 5900 & 9.86 \\ 
			SEQ2 & 5900 & 9.79 \\ 
			SEQ3 & 5900 & 9.65 \\ 
			\hline
	\end{tabular}}
\end{table}

In conclusion, Fast-MUOD provides a huge time performance gain over the original MUOD and Semifast-MUOD, despite its comparable or better outlier detection performance. Semifast-MUOD also provides some gains in running time over the original MUOD but not as much as Fast-MUOD, and its running time still increases with a factor dependent on $n^2$. All variants of MUOD used in these experiments were run using a single core. Since the MUOD methods have parallel implementations,  more performance gains can be obtained by running in parallel with more than one core, especially for Semifast-MUOD. 

\subsection{Sensitivity Analysis}
We evaluate the performance of all the methods with increased contamination rate of $\alpha = 0.15$ and $\alpha = 0.2$. The results are presented in Tables 2 and 3 of the Online Resource and they show the outlier detection results on Models 2 - 8. When $\alpha = 0.15$, the proposed methods maintain their good performance on Models 2 and 4 with slight reductions in outlier detection accuracy on Models 3, 5, 6, and 8. The most notable difference is the reduction in average TPRs of the proposed methods on Model 7 (e.g., FST reduces from $79.73\%$ to $41.90\%$) which is challenging as mentioned earlier (because the outliers are quite similar in shape and magnitude to the mass of the data).  Other competing methods suffer some reduction in performance also, including OGMBD on Models 3, 4, 5, and Model 8. The performance of TVD, FOM2 and FAO2 also reduces on Model 6 which contains pure shape outliers. At $\alpha = 0.2$, the proposed methods still maintain their performance on Models 2 and 4. There is a reduction in performance on Model 8 but the average TPRs of the proposed methods are still quite high at around $90\%$. The reduction in performance of the methods on Models 3, 5 and 6 (which all contain some form of shape outliers) are more pronounced. The proposed methods break down on Model 7, although other methods also break down on this model. OGMBD, TVD, FOM2 and FAO2 also reduce in outlier detection performance especially on the models with shape outliers. Worthy of mention is MSPLOT which maintains its outlier detection accuracy across the different contamination rates, except on Model 7 on which it did not perform well, even at $\alpha = 0.10$.
 
In Section 3 of the Online Resource, we evaluate the performance of the proposed methods on lower sample size of $n = 100$ and evaluation points of $d = 25$. The results of this experiment can be found in Table 4 of the Online Resource. Except in Model 7, the performance of the proposed models does not change much despite the reduction in sample size and number of evaluation points. In Model 7, there is a reduction in the accuracy of Fast-MUOD from an average TPR of $79.73\%$ (when $n =300$ and $d = 50$) to $73.42\%$. We also notice an increase in the standard deviation of the TPR from $14.95\%$ to $26.24\%$. This is an indication that the amplitude indices $I_A$ might be less sensitive to outliers with reduced sample size or evaluation points. It is also worthy of note that the TPR of other competing methods like MSPLOT and TVD decreased from $66.39\%$ and $40.88\%$ respectively to $48.42\%$ and $27.38\%$ respectively. The standard deviations of their TPR also increased from $16.04\%$ and $12.82\%$ respectively to $27.14\%$ and $24.90\%$ respectively in this model.

Section 5 of the Online Resource shows the performance of the proposed methods when the signal-to-noise ratio in the simulated data is increased or decreased. We do this by increasing or decreasing the variance of the simulation models. Using Models 2, 3, 4, and 6, we change the covariance matrix in the base and contamination models to $\gamma(s, t) = \nu \cdot \exp{-|t - s|}$, where $s, t \in [0, 1]$ and $\nu \in \{0.25, 0.5, 1.5, 5\}$. At lower variance levels ($\nu \in \{0.25, 0.5\}$), the TPRs of the proposed methods increased, especially on Model 6, because of the reduced noise in the data. However, at higher variance levels $(\nu \in {1.5, 5})$, the proposed methods starts to break down due to the increased noise. This breakdown is also seen in other competing outlier detection methods. When $\nu  = 1.5$, all the methods still perform well on Model 2 with magnitude outliers and our proposed
methods still maintain a good performance for Model 4. When $\nu = 5$ though, all the methods break down with low TPRs on all the models considered except for TVD on Model 3, FOM2 and FAO2 on Models 2 and 3 and SEQ1 and SEQ3 on Model 3. We refer the reader to Tables 6 and 7 of the Online Resource for detailed results of the experiment.

\section{Applications}
\label{sec::app}
In this section, we apply the Fast-MUOD method to three scenarios: outlier detection in weather data, object recognition in surveillance video data, and population growth patterns of countries.

\subsection{Spanish Weather Data}
\begin{figure*}[htbp!]
	\centering
	\includegraphics[width=1\linewidth]{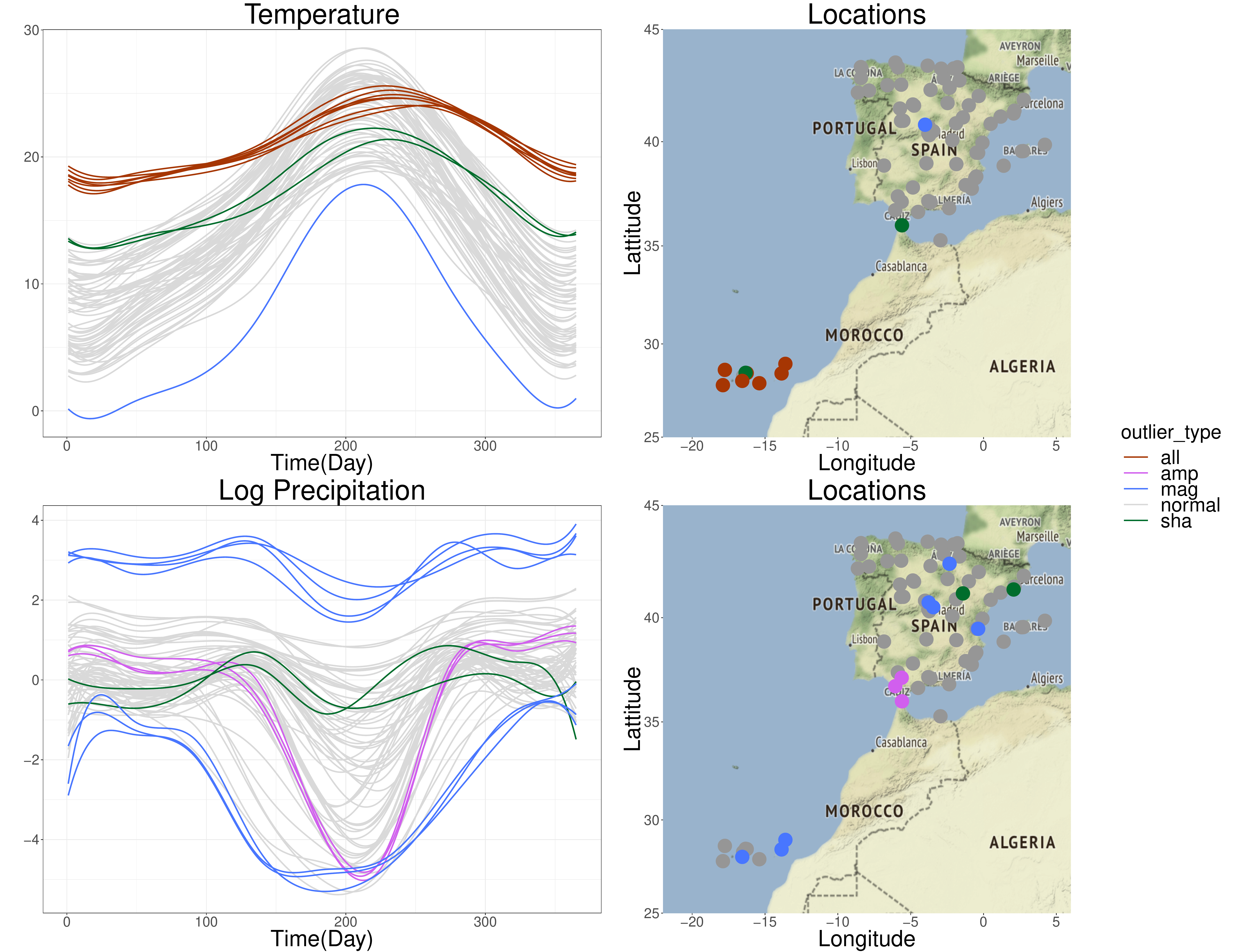} 
	\caption{\label{fig:aemet_fast_type} Curves flagged as outliers by Fast-MUOD. First Column: smoothed Temperature curves (top), and smoothed Log Precipitation curves (bottom). Second Column: geolocations of weather stations. Legend: curves flagged as magnitude, amplitude and shape outliers (all, in orange), curves flagged as magnitude outliers only (mag, in blue), curves flagged as shape outliers only (sha, in green), curves flagged as amplitude outliers only (amp, in purple), non-outlying curves (normal, in gray).}
\end{figure*}

The Spanish weather data collected by the ``Agencia de Estatal de Meteorologia" (AEMET) of Spain, contains daily average temperature, precipitation, and wind speed of 73 Spanish weather stations between the period 1980-2009. Geographical information about the location of these stations are also recorded in the data. This dataset is available in the \textit{fda.usc}  (\citeauthor{fda.usc} \citeyear{fda.usc}) R package and it has been analysed in FDA literature, e.g., \cite{dai2018multivariate}. For this analysis, we use the temperature and log precipitation data. As done in \cite{dai2018multivariate}, we first smooth the data, we then run Fast-MUOD on the smoothed data and collate the different types of outliers flagged for both temperature and log precipitation. The first column of Figure \ref{fig:aemet_fast_type} shows the different outliers flagged while the second column shows the geographical locations of the flagged outliers.

For temperature, seven weather stations on the Canary Islands are flagged simultaneously as amplitude, shape and magnitude outliers because of the different prevailing weather conditions on this archipelago compared to the other stations located in mainland Spain. Furthermore, two pure shape outliers are flagged, one located on the Canary Islands and the other on the southern tip of Spain, close to the Strait of Gibraltar. The temperature in these regions changes more gradually over the year than in mainland Spain. Finally, a single magnitude outlier is flagged, albeit a lower magnitude one. This weather station records lower temperatures all through the year compared to the other stations because it is located at a very high altitude in the ``Puerto de Navacerrada" mountain pass in the north of Madrid. This station has the highest altitude of all the weather stations in mainland Spain and is known to experience cold temperatures. 

For log precipitation, two groups of magnitude outliers are identified, with the first group (of four stations) recording higher precipitation on the average. The second group of three stations are located on the Canary Islands where it is dryer on the average all through the year. A group of pure amplitude outliers, containing 3 stations, is also flagged by Fast-MUOD. These stations experience a more abrupt decline in precipitation during the summer months compared to the more gradual decline in precipitation experienced in other stations located in Spain's interior. These three stations are located in the southern tip of Spain which is known to experience dry summer months. Finally a cluster of pure shape outliers made up of two stations is flagged. The curves of these two stations seem to vary more through the year. One of these station is located in Barcelona, on the eastern coast of Spain which is known to be humid and rainy. The other station is located in Zaragoza, with wet periods during the spring and autumn months.

\begin{figure*}[htb!]
	\centering
	\includegraphics[width=1\linewidth]{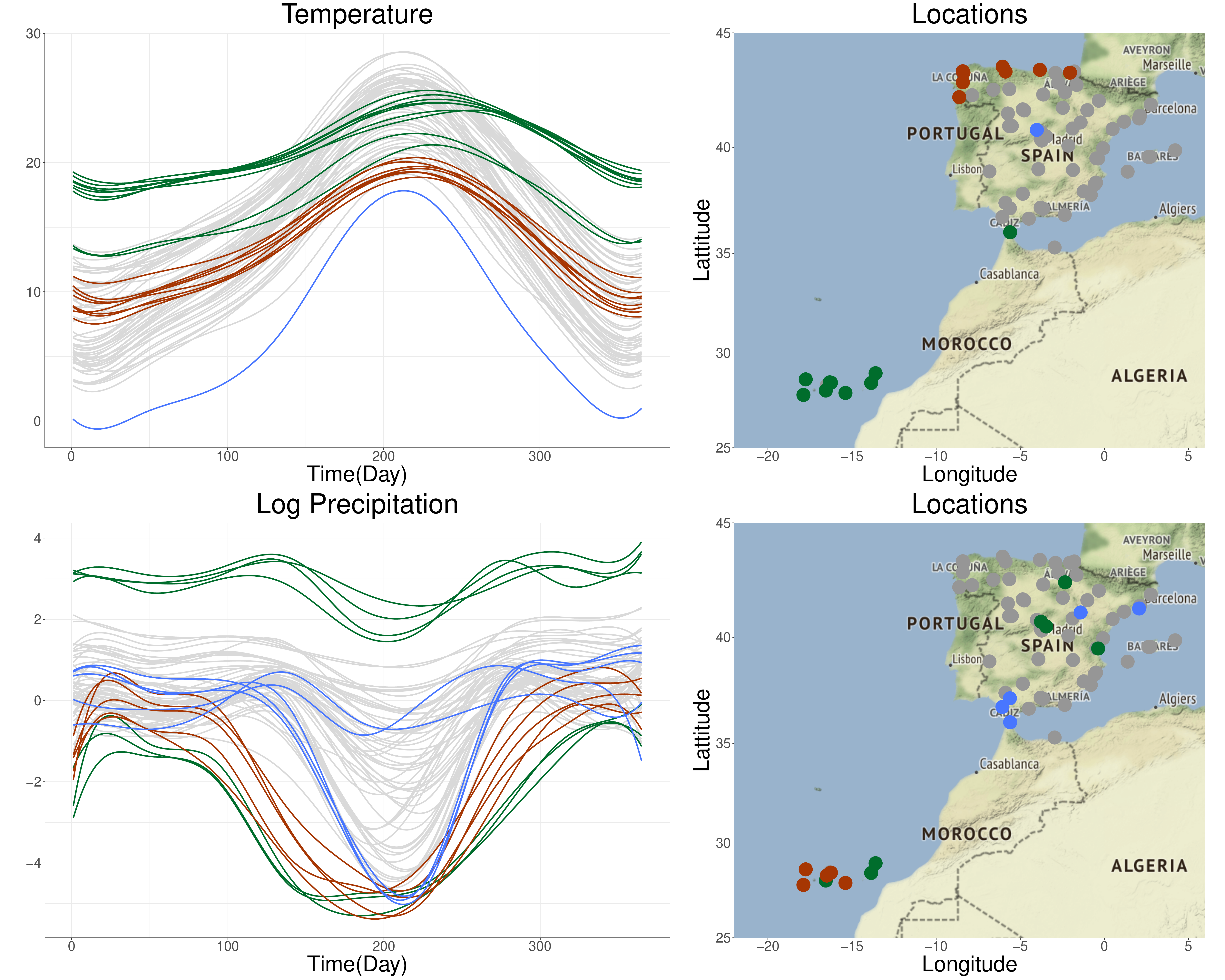} 
	\caption{Curves flagged as outliers by Fast-MUOD and MS-plot. First column: smoothed Temperature curves (top), and smoothed Log Precipitation curves (bottom). Second column: Geolocations of weather stations.
		Color code: Curves flagged as outliers by Fast-MUOD and MS-plot (green), curves flagged as outliers by MS-plot only (orange), curves flagged as outliers by Fast-MUOD only (blue).\label{fig:aemet_fast_msplot} }
\end{figure*}

We compared the results of our analysis to those of MS-plot obtained by \cite{dai2018multivariate}. Even though MS-plot is for multivariate functional data visualization and outlier detection, we chose it because it also handles univariate outlier detection  quite well as shown by the results of our simulation studies. In doing this comparison, we combined all the outliers of different types flagged by Fast-MUOD and compared them to those flagged by MS-plot. Figure \ref{fig:aemet_fast_msplot} shows the results of both methods.

For temperature, Fast-MUOD and MS-plot both flag as outliers all the weather stations on the Canary Islands and the single station in the south tip of Spain by the Strait of Gibraltar. However, only MS-plot flags the stations in the north of Spain as outliers, while only Fast-MUOD flags as outlier the single lone station in Madrid where significantly lower temperatures are recorded all through the year.  Likewise, for log precipitation, both methods flag as outliers the four stations with significantly higher precipitation than the remaining stations. MS-plot flags as outliers all the stations in the southern Canary Islands, while Fast-MUOD flags only some of them, specifically those with the lowest precipitation for most part of the year. Only Fast-MUOD flags as outliers the three stations in the southern tip of Spain where there is a sharper decline in precipitation during the summer months, because of its ability to detect amplitude outliers. Furthermore, two additional stations in Barcelona and Zaragoza, flagged by Fast-MUOD as shape outliers, were not flagged by MS-plot. Even though there is no ground truth as to which stations are outliers or not in this dataset, both Fast-MUOD and MS-plot flagged reasonable outliers and the classification of outliers into types by Fast-MUOD could be an advantage since it is not necessary to visualize the data to know why an observation is an outlier.


\subsection{Surveillance Video}
\label{subsec:surveillance}
Next we apply Fast-MUOD on a surveillance video data named \textit{WalkByShop1front}. This video was filmed by a camera across the hallway in a shopping centre in Lisbon. The video is made available online by the CAVIAR project at \href{http://homepages.inf.ed.ac.uk/rbf/CAVIARDATA1/}{homepages.inf.ed.ac.uk/rbf/CAVIARDATA1}. The video shows the front of a clothing store with people walking through the corridor in front of the shop. The video is about 94 seconds long and at different times in the course of the video, five people passed by the front of the shop, two of whom entered the store to check clothes in the store. The objective is to use Fast-MUOD to identify points in the video when people passed by the front of the shop. 

With each second of the video consisting of 25 frames, the video contains a total of 2359 frames. Each of the frame is made up of $384\times 288 = 110592$ pixels. We first convert the RGB values of the pixels of the each frame to gray scale. From the matrix of gray intensity values of each frame, we form a row vector of length $110592$ by ordering the values of the matrix column-wise (we obtained the same result by arranging row-wise). The constructed functional data is made up of 2359 curves observed at $110592$ points. 

\begin{figure*}[htbp!]
	\centering
	\includegraphics[width=1.04\linewidth]{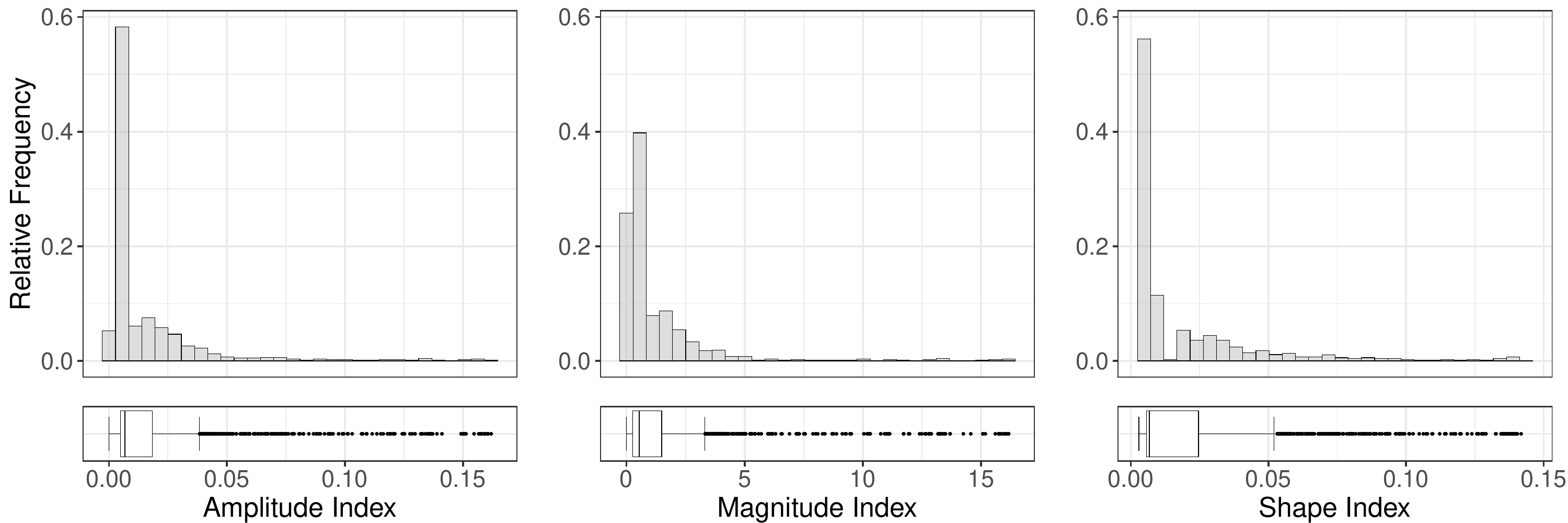} 
	\caption{Distribution of the amplitude, shape and magnitude indices of the video data.
		\label{fig:index_dist} }
\end{figure*}
\begin{figure*}[htbp!]
	\centering
	\includegraphics[width=1\linewidth]{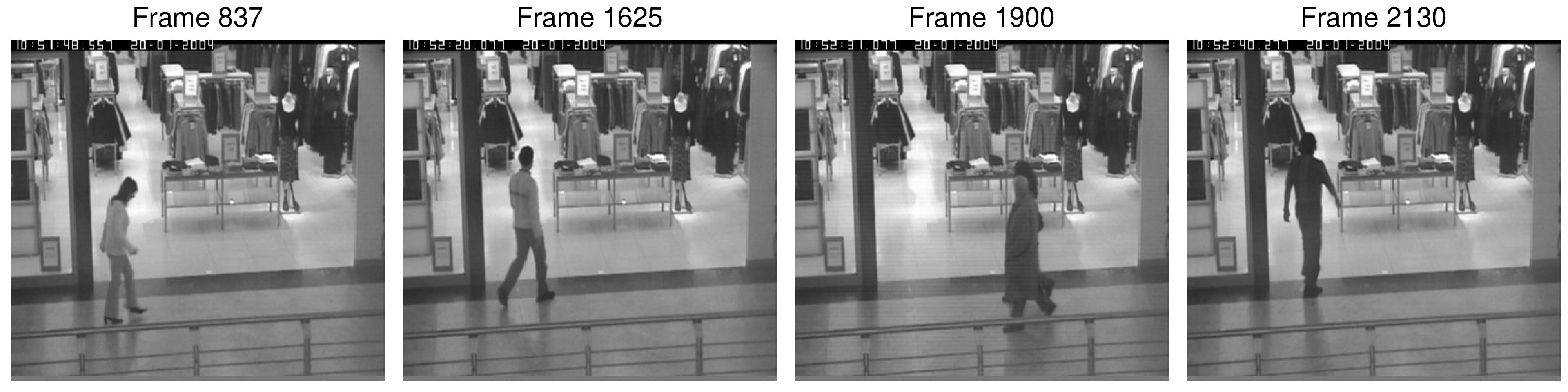} 
	\caption{Some outliers detected by the Fast-MUOD from the video.}
	\label{fig:vid_detect}
	
\end{figure*}
We then apply Fast-MUOD on the constructed functional data using the point-wise median to speed up computation. Figure \ref{fig:index_dist}  shows the histogram and the boxplot of magnitude, amplitude and shape indices from this data. We obtained 216 shape outliers, 206 amplitude outliers and 194 magnitude outliers. The three types of outliers flagged are not mutually exclusive as shape outliers for instance can also have partial magnitude outlyingness. There are 125 outliers that are outliers of the three types, i.e., they are flagged simultaneously as magnitude, amplitude and shape outliers. There are only 34 pure magnitude outliers, 15 pure amplitude outliers and 48 pure shape outliers. In total, there are 294 unique frames flagged as outliers.

\begin{figure*}[htbp!]
	\centering
	\includegraphics[width=1\linewidth]{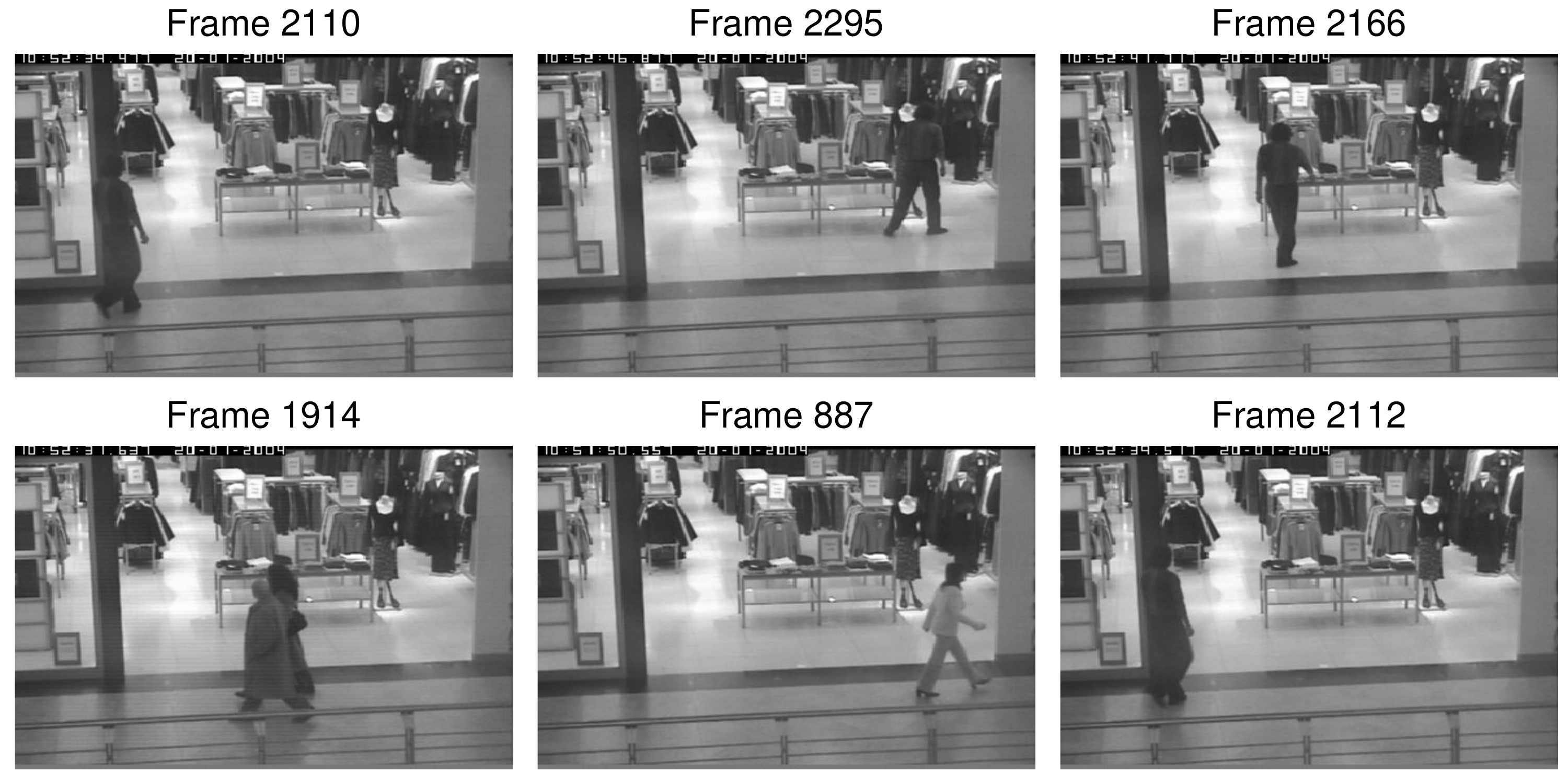} 
	\caption{Selected frames from the video in gray scale. Frames 2110 and 2295: frames not detected as outliers. Frame 2166: sample pure magnitude outlier. Frame 1914: sample pure shape outlier. Frame 887 and 2112: sample pure amplitude outliers. 
		\label{fig:vid_nodetect} }
\end{figure*}

All the 294 outlying frames correspond to time points in the video when people passed by the front of the store, thus there are no false positives. For instance, frames 831 - 846 and 885 - 887 correspond to the period when the first person in the video passed by the front of the store (see Figure \ref{fig:vid_detect}, Frame 837). The same for frames 1614 - 1642 when the second person passed by and entered the shop (Figure \ref{fig:vid_detect}, Frame 1625). Frames 1852 - 1983 correspond to when two women passed by together in front of the shop (Figure \ref{fig:vid_detect}, Frame 1900) and frames 2112 - 2169 and 2296 - 2336 which correspond to the period when the last person passed by and entered the shop (Figure \ref{fig:vid_detect}, Frame 2130). There are small pockets of time periods (frames) in the video (which are in between the frames flagged as outliers), that contain people but are not flagged as outliers. We notice that these usually corresponds to time periods where there is not a enough contrast (from the gray intensities) between the person in the video and the environment because we converted the frames to gray scale before analysis. This is seen in Frame 2110 (Figure \ref{fig:vid_nodetect}) for instance, when a man wearing a dark blue and red shirt with dark trousers was standing entirely behind a dark pillar or in Frame 2295 (Figure \ref{fig:vid_nodetect}) when the same man was standing beside dark clothes in the store.

Examining the outliers of each type also provide some additional insight. The 34 pure magnitude outliers are frames which contain the man wearing a dark clothes as he entered into the shop, e.g., Frame 2166 (Figure \ref{fig:vid_nodetect}). The gray intensities (and hence contrast with the environment) at this time is very high due to the dark nature of his clothing and hence the reason why these frames are pure magnitude outliers. The 48 pure shape outliers correspond to frames that contain the two women passing by together in front of the shop, e.g., Frame 1914 in Figure \ref{fig:vid_nodetect}. The 15 pure amplitude outliers are frames where people just entered or are about to exit the field of view of the camera (see Frames 887 and 2112). Thus, in addition to detecting outlying frames, the classification of the different frames also give some insight which might prove valuable in different use cases. 

\cite{huang2019decomposition} and \cite{rousseeuw2018measure} applied their methods on video data. For comparison, we run their methods (TVD and FOM) and MSPLT on the constructed functional data obtained from the surveilance video. While MSPLT did not produce any result due to some computational error, TVD flagged all the frames as outliers after running for over 9 hours (compared to about 42 seconds of Fast-MUOD). FOM however performed excellently, flagging only frames that include people passing by the shop and producing the result in a reasonable time of  193 seconds. 

\subsection{Population Data}
Finally, we analyse the world population data from the United Nations, also analysed by \cite{nagy07depth} and \cite{dai2020sequential}. This data contains the yearly total population of 233 countries (and autonomous regions) recorded in the month of July, 1950 to 2015. Following \cite{nagy07depth} and \cite{dai2020sequential}, we select only countries with population between one million and fifteen million of 1980 which leaves us with 105 countries out of the total 233 countries. The constructed functional data is then made up of 105 curves observed at 65 points. We apply Fast-MUOD and the countries/regions detected as outliers are shown in Table \ref{tab:fastmuod-country}.

\begin{table*}[t]
	\caption{Countries detected as outliers by Fast-MUOD \label{tab:fastmuod-country}}
	\centering
	\setlength{\tabcolsep}{5pt}

	{\renewcommand{\arraystretch}{.9}
		\normalsize
		\begin{tabular}{P{4cm}P{4cm}P{4cm}}
			\hline
			Magnitude outliers & Amplitude outliers & Shape outliers  \\
			\hline
			Saudi Arabia, Sudan, Uganda & Sudan, Uganda, Saudi Arabia, Iraq, Malaysia, Yemen, Afghanistan,
			Ghana, Nepal, C\^ote d'Ivoire, Mozambique, Madagascar, Angola, Syrian Arab Republic, Cameroon &  Bulgaria, Latvia, Hungary, Georgia, Croatia, Estonia, Lithuania, Bosnia and Herzegovina, Belarus, Armenia, Serbia, Republic of Moldova, Kazakhstan, Albania, Czech Republic, United Arab Emirates, TFYR Macedonia, Slovakia\\
			\hline
	\end{tabular}}
\end{table*}

\begin{figure*}[htbp!]
	\centering
	\includegraphics[width=.9\linewidth]{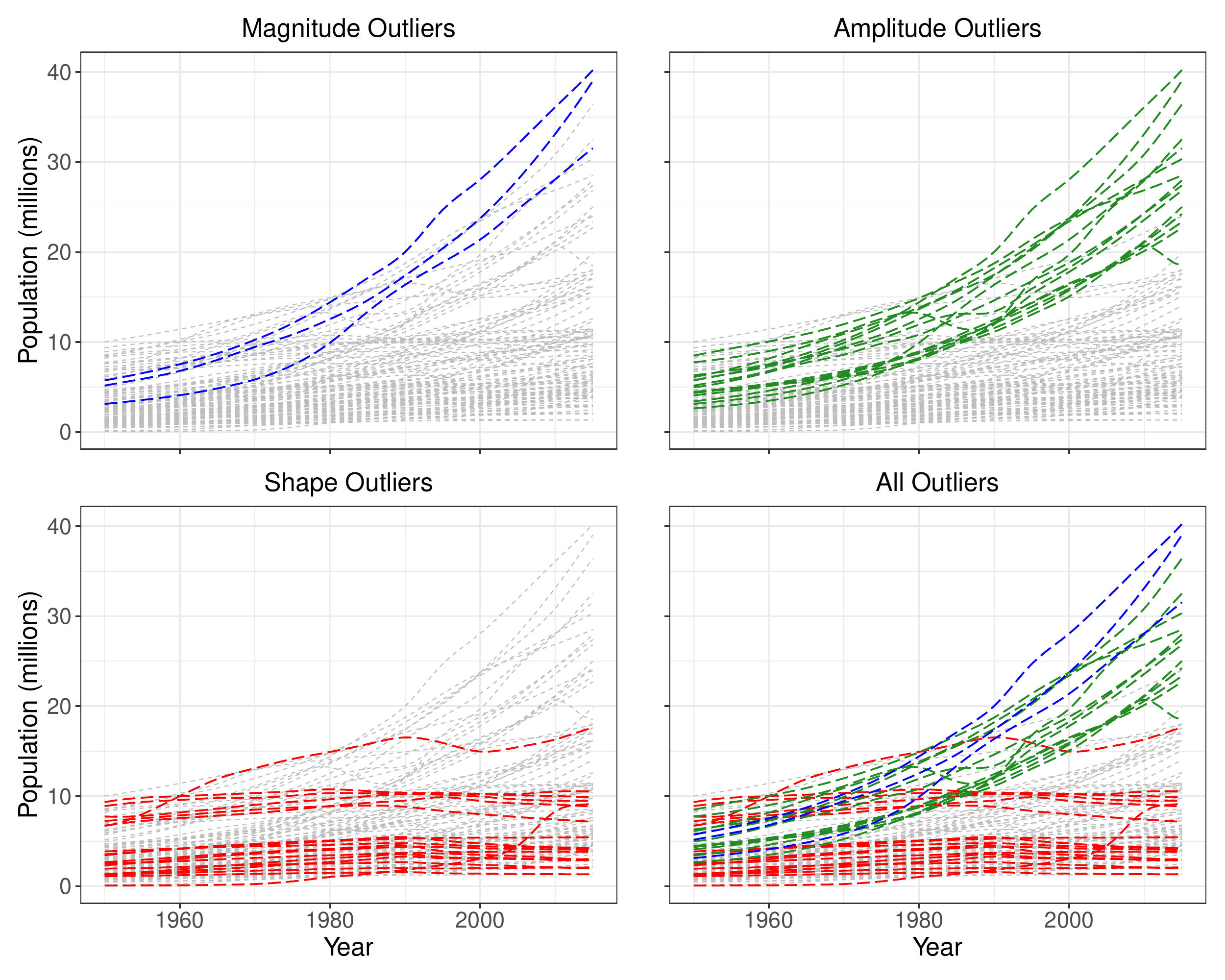} 
	\caption{Outliers detected by Fast-MUOD from the population data. Top-left: Magnitude outliers. Top-right: Amplitude outliers. Bottom-left: Shape outliers. Bottom-right: All the outliers. }
	\label{fig:population}
\end{figure*}

In total there are 33 unique countries detected as outliers; 3 of them magnitude outliers, 15 of them amplitude outliers and 18 of them shape outliers. Again, the types of outliers are not mutually exclusive as all magnitude outliers are also amplitude outliers. Saudi Arabia, Sudan and Uganda are flagged as magnitude outliers because they had highest population values toward the end of the period of the data (2015), as can be seen in Figure \ref{fig:population}. Sudan, despite having a population of about 5 million in 1950, had the highest population value of 40 million 2015. The same trend is observed for Uganda and Saudi Arabia, with population of 39 million and 31 million respectively in 2015. The amplitude outliers are shown in the top-right panel of Figure \ref{fig:population}. These are countries with very high population growth rate in the period of the data. Among these are Sudan, Uganda and Iraq, as they had an increase of 34 million, 33 million and 30 million respectively between 1950 and 2015. Other countries include Saudi Arabia, Afghanistan and Malaysia. All the countries flagged as amplitude outliers are either located in the Middle East or Africa.  

Finally, the shape outliers are shown in the bottom-left panel of Figure \ref{fig:population}. The curves of these countries show a different shape and trend compared to the other countries. One observation about these countries is their peculiar pattern of a slight increase in population growth till 1980 followed by a plateau or slight decrease in the population till the end of the study period. There are also few countries with a sharp increase or decrease in population. Furthermore, all these countries except for United Arab Emirates (UAE) are located in Central and Eastern Europe with similar demographics, geographical, economic and political environment. 

Compared to the results obtained in \cite{nagy07depth}, our method identifies more outliers. For instance, the first order outliers (which are equivalent to magnitude outliers) identified in \cite{nagy07depth} did not include Sudan which had the highest population by the end of the investigated period (2010). Though Sudan  was flagged as a second order outliers, a lot of countries in Eastern Europe flagged as shape outliers were not flagged as outliers. Except for Netherlands, all the second and third order outliers flagged by \cite{nagy07depth} are also flagged as either shape outliers or amplitude outliers by Fast-MUOD.  Furthermore, our classification of the different outliers provide additional information and consistent interpretation on why observations are flagged as outliers.

\cite{dai2020sequential} also analysed this data and classified outliers found using sequential transformations. All the ``pattern" outliers found are countries in Eastern Europe except for Rwanda just like the shape outliers flagged by Fast-MUOD. In fact, our method flagged all the pattern outliers found by \cite{dai2020sequential} as shape outliers except for Rwanda. However, Fast-MUOD flags five additional shape outliers including Macedonia, Serbia, Albania, and Slovakia, all located in Central and Eastern Europe. The amplitude outliers flagged by Fast-MUOD also include all the amplitude outliers flagged by the method described in \cite{dai2020sequential} and all these countries are located in the Middle East and Africa too. While Fast-MUOD flagged only three magnitude outliers compared to nine magnitude outliers flagged by \cite{dai2020sequential}, the remaining six magnitude outliers were flagged by Fast-MUOD as amplitude outliers. In fact, some of these six magnitude outliers were flagged as  both magnitude and amplitude outliers by \cite{dai2020sequential}, but they were grouped as magnitude outliers in order to maintain a mutually exclusive classification of outliers (see Table 5 in \cite{dai2020sequential} for details).

Overall, our results are quite consistent with those obtained by \cite{dai2020sequential} even though there are slight differences in classification of some outliers. Also worthy of note is the slight difference in investigated period (1950 - 2010) in the analysis by \cite{dai2020sequential} and \cite{nagy07depth}.

\section{Discussion} 
\label{sec:conc} 

In this paper, we have proposed two methods based on the MUOD outlier detection method. These methods use a sample of the data to compute the indices for Semifast-MOUD, or a median ($L_1$ or point-wise) in the case of Fast-MUOD and they improve on the scalability and outlier detection performance of MUOD. In separating the outlier indices from the indices of the typical observations, we use the classical boxplot. All these put together make the proposed methods intuitive and based on simple statistical concepts, consequently making them less computationally intensive. Different types of outliers are identified and classified directly, giving an intuition as to why a curve is flagged as an outlier without the need for visualization or manual inspection. This is valuable in cases where manual inspection or visualizing the data is difficult. 

Using both simulated and real data, we have shown the performance benefits of these methods over MUOD. Further comparisons to existing univariate functional outlier detection tools shows comparable or superior results in correctly identifying potential outliers of different types. Implementation is done in R and the code is made available at \href{https://github.com/otsegun/fastmuod}{https://github.com/otsegun/fastmuod}. 

Possible further improvement could be extension of the methods to multivariate functional data. The use of orthorgonal regression in the computation of the indices is interesting to study. Exploring the theoretical properties of the MUOD indices is also a possible next line of investigation.

\section*{Supplementary Materials}

\begin{description}

	\item[Simulation Results:] Additional simulation results showing comparisons between Fast-MOUD computed with the $L_1$ median and the point-wise median and outlier detection performance of all methods considered at higher contamination rates.  Also includes comparisons of different correlation coefficients for computing the shape index $I_S$. More simulations results using lower sample size and lower evaluation points are also presented together with a sensitivity analysis of outlier detection performance when more noise is added to the simulation models.

\end{description}

\bibliographystyle{ECA_jasa}
\bibliography{bib_full}

\newpage
\section*{Suplementary Material: Detecting and Classifying Outliers in Big Functional Data}
\subsection*{Comparison between $L_1$ median and Point-wise median for Fast-MUOD}
In this section, we present simulation results showing that the performance of Fast-MUOD using the point-wise median (FSTP) and the $L_1$ median (FSTL1) are similar. Note that FSTL1MAG considers magnitude outliers only, flagged by the magnitude index of FSTL1. The same applies to FSTL1SHA and FSTL1AMP (considering shape outliers and amplitude outliers only of FSTL1 respectively). Thus, FSTL1 is the union of outliers flagged by FSTL1MAG, FSTL1SHA, and FSTL1AMP. The same notation system is used for FSTP.  
The results can be found in Table \ref{tab:sim-l1-pwise}.
\label{sec:pointwisel1}

\subsection*{Contamination rate}
Here, we present simulation results showing that the performance of the proposed methods as the contamination rate is increased up to $20\%$, Results are presented in Tables \ref{tab:sim-results-300-50-0.15} and \ref{tab:sim-results-300-50-0.2}.
\label{sec:contamination_rate}

\subsection*{Sample Size and Evaluation Points}
Here, we present simulation results showing that the performance of the proposed methods, Semifast-MUOD and Fast-MUOD, remain mostly similar, except for Model 7,  even with lower sample size and evaluation points of $n = 100$ and $d = 25$ respectively. See Subsection 4.2 of the main manuscript for a description of the methods compared. Results are presented in Table \ref{tab:sim-results-100-25}.
\label{sec:more res}

\subsection*{Correlation coefficients}
We compare the effectiveness of different correlation coefficients that can be used in computing the shape index $I_S(Y_i, \tilde{Y})$ for Fast-MUOD.  We considered Pearson, Spearman's rank, and Kendall's Tau correlation coefficients, in addition to the Cosine similarity index. We used Models 3, 4, and 6 all which have some form of shape outliers. We then recorded the True Positive Rate (TPR) and the False Positive Rate (FPR) (together with their standard deviations) of the outliers flagged by only the shape indices. Specifically, the methods considered are: 
\begin{itemize}
    \item FSTSH\_PEARSON: An observation is an outlier only if it is flagged by Fast-MUOD as a shape outlier using $I_S(Y_i, \Tilde{Y})$ computed with the Pearson's correlation coefficient.
    \item FSTSH\_KENDALL: An observation is an outlier only if it is flagged by Fast-MUOD as a shape outlier using $I_S(Y_i, \Tilde{Y})$ computed with the Kendall's Tau correlation coefficient.
   \item FSTSH\_SPEARMAN: An observation is an outlier only if it is flagged by Fast-MUOD as a shape outlier using $I_S(Y_i, \Tilde{Y})$ computed with Spearman's rank correlation coefficient. 
   \item FSTSH\_COSINE: An observation is an outlier only if it is flagged by Fast-MUOD as a shape outlier using $I_S(Y_i, \Tilde{Y})$ computed with the Cosine similarity index.
\end{itemize}

Table \ref{tab:sim-results-correlation} shows the results of our simulation. The shape index $I_S(Y_i, \tilde{Y})$ computed using the Pearson correlation coefficient identifies more shape outliers, especially in Model 3 and Model 6.

\subsection*{Signal to Noise Ratio}
\label{sec:variance}
Here we study the changes in the  True Positive Rate (TPR) and the False Positive Rate (FPR) while considering different level of noise in the simulated data. To do this, we change the covariance matrix in the base and contamination models for Models 2, 3, 4, and 6 to $\gamma(s,t) = \nu\cdot \exp\{-|t-s|\}$, where $s,\ t \in [0, 1]$ and $\nu \in \{0.25, 0.5, 1.5, 5\}$. At lower variance levels ($\nu \in \{0.25, 0.5\}$), the proposed methods perform better (compared to when $\nu = 1$) with improved TPR, especially on Model 6, because of the reduced noise in the data. However, at higher variance levels ($\nu \in \{ 1.5, 5\}$), the proposed methods starts to breakdown, just like the other methods, due to the increased noise. At $\nu = 1.5$, all the methods still perform well on Model 2 with magnitude outliers and our proposed methods maintain a good performance for Model 4. At $\nu = 5$, all the methods breakdown with low TPRs on all the models except for TVD on Model 3. These results can be seen in Tables \ref{tab:snr1} and \ref{tab:snr2}.

\begin{table*}[h!]
\caption{\label{tab:sim-l1-pwise}Mean and Standard Deviation (in parentheses) of the True Positive Rate (TPR) and the False Positive Rate (FPR) over eight simulation models comparing the point-wise median and the $L_1$ median for computing the Fast-MUOD Indices. Experiment setup include 500 repetitions with $n=300$, $d=50$, and $\alpha = 0.1$.}
\centering
\renewcommand{\arraystretch}{.7}
{
  \footnotesize
\begin{tabular}{@{}lccccccc@{}}  \toprule
\multirow{2}{*}{Method} & Model1  & \multicolumn{2}{c}{Model 2} & \multicolumn{2}{c}{Model 3} & \multicolumn{2}{c}{Model 4} \\
\cmidrule{2-2} \cmidrule{3-4} \cmidrule{5-6} \cmidrule{7-8}
& FPR & TPR & FPR & TPR & FPR & TPR & FPR \\ 
\hline
FSTL1 & 9.91(1.53) & 100.00(0.00) & 8.94(1.53) & 99.79(0.98) & 6.10(1.37) & 100.00(0.00) & 3.15(1.10) \\ 
 FSTL1MAG & 1.72(0.92) & 99.99(0.15) & 0.36(0.39) & 4.23(3.66) & 1.52(0.89) & 41.99(9.76) & 0.62(0.53) \\ 
 FSTL1SHA & 7.95(1.37) & 7.74(4.85) & 7.94(1.43) & 98.98(2.01) & 4.35(1.14) & 100.00(0.00) & 2.25(0.94) \\ 
 FSTL1AMP & 1.71(0.85) & 1.67(2.39) & 1.69(0.93) & 6.39(4.70) & 1.39(0.79) & 55.74(11.72) & 0.45(0.45) \\ 
FSTP & 9.90(1.50) & 100.00(0.00) & 8.95(1.59) & 99.81(0.89) & 6.10(1.37) & 100.00(0.00) & 3.15(1.13) \\ 
FSTPMAG & 1.74(0.92) & 99.99(0.15) & 0.36(0.40) & 4.13(3.70) & 1.52(0.89) & 41.25(9.71) & 0.63(0.54) \\ 
 FSTPSHA & 7.94(1.34) & 7.79(4.86) & 7.94(1.48) & 98.97(2.03) & 4.36(1.12) & 100.00(0.00) & 2.24(0.95) \\ 
FSTPAMP & 1.70(0.83) & 1.66(2.43) & 1.69(0.93) & 6.41(4.62) & 1.38(0.80) & 54.56(11.76) & 0.45(0.45) \\ 

\end{tabular}}
\setlength\tabcolsep{3.0pt} 
{\renewcommand{\arraystretch}{.7}
  \footnotesize
\begin{tabular}{@{}lcccccccc@{}}  \toprule
\multirow{2}{*}{Method} & \multicolumn{2}{c}{Model 5} & \multicolumn{2}{c}{Model 6} & \multicolumn{2}{c}{Model 7} & \multicolumn{2}{c}{Model 8} \\
  \cline{2-3} \cline{4-5} \cline{6-7} \cline{8-9}
& TPR & FPR & TPR & FPR & TPR & FPR & TPR & FPR\\ 
  \hline
FSTL1 & 96.11(4.22) & 5.67(1.18) & 93.31(6.25) & 6.33(1.37) & 79.69(15.05) & 6.56(1.89) & 98.75(2.28) & 6.64(1.41) \\ 
FSTL1MAG & 16.02(6.73) & 1.08(0.71) & 0.81(1.65) & 1.76(0.96) & 1.64(2.34) & 1.69(0.89) & 30.65(8.12) & 1.05(0.75) \\ 
FSTL1SHA & 86.50(6.79) & 4.40(1.11) & 91.26(6.67) & 4.37(1.13) & 4.23(3.68) & 4.95(1.69) & 71.92(7.60) & 5.31(1.29) \\ 
FSTL1AMP & 23.10(7.96) & 1.02(0.72) & 3.53(3.60) & 1.42(0.79) & 79.03(15.59) & 0.01(0.05) & 10.71(5.72) & 1.29(0.84) \\ 
FSTP & 95.97(4.27) & 5.67(1.19) & 93.05(6.42) & 6.31(1.35) & 79.73(14.95) & 6.55(1.91) & 98.63(2.45) & 6.65(1.40) \\ 
FSTPMAG & 15.94(6.70) & 1.08(0.71) & 0.83(1.70) & 1.77(0.94) & 1.65(2.36) & 1.69(0.90) & 30.65(8.10) & 1.04(0.75) \\
FSTPSHA & 86.35(6.74) & 4.39(1.12) & 91.01(6.75) & 4.35(1.10) & 4.21(3.68) & 4.94(1.72) & 71.77(7.58) & 5.31(1.29) \\ 
FSTPAMP & 22.99(7.93) & 1.01(0.71) & 3.54(3.78) & 1.40(0.79) & 79.10(15.42) & 0.01(0.05) & 10.74(5.69) & 1.29(0.83) \\ 
\bottomrule
\end{tabular}}
\end{table*}

\begin{table*}[h!]
\centering
\caption{\label{tab:sim-results-300-50-0.15}Mean and Standard Deviation (in parentheses) of the True Positive Rates (TPR) and False Positive Rate (FPR) over eight simulation models with 500 repetitions for each possible case. Each simulation is done with $n=300$ and $d=50$ and $\alpha = 0.15$. Comparatively high TPRs are in bold. Proposed methods in italics.}	
{\renewcommand{\arraystretch}{.7}
\scriptsize
\begin{tabular}{@{}lllllllll@{}} 
\toprule
\multirow{2}{*}{Method} & \multicolumn{2}{c}{Model 2}  & \multicolumn{2}{c}{Model 3} & \multicolumn{2}{c}{Model 4} & \multicolumn{2}{c}{Model 5} \\
\cmidrule{2-3} \cmidrule{4-5} \cmidrule{6-7} \cmidrule{8-9}

& TPR & FPR & TPR & FPR & TPR & FPR & TPR & FPR \\ 
 \midrule
\textit{FST}   & \textbf{100.00(0.00)} & 8.89(1.58) & \textbf{96.48(4.37)} & 4.31(1.14) & \textbf{99.98(0.20)} & 1.38(0.77) & \textbf{88.65(7.05)} & 3.82(1.13) \\ 
\textit{FSTMG}  & \textbf{100.00(0.00)} & 0.09(0.20) & 3.97(2.95) & 1.40(0.84) & 30.94(7.84) & 0.34(0.40) & 14.82(5.42) & 0.91(0.66) \\ 
\textit{FSTSH} & 7.98(4.20) & 8.02(1.55) & \textbf{94.01(5.36)} & 2.65(0.87) & \textbf{99.98(0.22)} & 0.94(0.63) & 77.31(7.95) & 2.78(0.95) \\ 
\textit{FSTAM}  & 1.79(1.92) & 1.69(0.97) & 6.27(3.55) & 1.22(0.76) & 31.72(10.31) & 0.14(0.24) & 21.01(6.41) & 0.72(0.60) \\ 
\textit{SF} &  \textbf{99.99(0.17)} & 8.57(1.60) & \textbf{95.25(4.98)} & 3.91(1.16) & \textbf{98.91(1.85)} & 0.95(0.66) & 84.76(7.43) & 3.52(1.08) \\ 
\textit{SF25} & \textbf{99.96(0.42)} & 8.55(1.60) & \textbf{95.14(4.97)} & 3.86(1.11) & \textbf{98.26(3.19)} & 0.94(0.66) & 84.57(7.52) & 3.52(1.06) \\ 
\textit{MUOD}  & \textbf{99.62(4.90)} & 8.18(3.92) & 48.67(23.19) & 9.87(4.57) & 88.41(15.03) & 3.77(3.50) & 42.63(12.84) & 3.52(2.70) \\ 
OGMBD  & \textbf{100.00(0.10)} & 4.72(1.46) & 29.70(9.91) & 2.97(1.06) & 58.88(12.56) & 0.50(0.45) & \textbf{89.67(5.40)} & 0.94(0.66) \\ 
MSPLT  & \textbf{99.93(0.39)} & 2.39(1.21) & \textbf{100.00(0.00)} & 2.47(1.16) & \textbf{99.80(0.70)} & 1.15(0.77) & \textbf{99.98(0.20)} & 2.39(1.17) \\ 
TVD  & \textbf{99.99(0.14)} & 0.00(0.02) & \textbf{100.00(0.00)} & 0.00(0.00) & 1.48(2.09) & 0.00(0.00) & \textbf{100.00(0.00)} & 0.00(0.02) \\ 
FOM  & \textbf{100.00(0.00)} & 0.01(0.07) & 21.02(14.78) & 0.02(0.09) & 6.06(5.86) & 0.01(0.05) & 4.70(4.43) & 0.01(0.07) \\ 
FAO  & \textbf{100.00(0.10)} & 0.00(0.03) & 8.56(8.91) & 0.00(0.04) & 0.07(0.38) & 0.00(0.04) & 2.52(3.14) & 0.00(0.03) \\ 
FOM2  & \textbf{100.00(0.00)} & 0.51(0.47) & \textbf{100.00(0.00)} & 1.23(0.74) & 34.47(10.73) & 0.49(0.47) & \textbf{100.00(0.00)} & 1.16(0.73) \\ 
FAO2  & \textbf{100.00(0.00)} & 0.70(0.56) & \textbf{100.00(0.00)} & 0.89(0.67) & 4.27(3.47) & 1.39(0.87) & \textbf{100.00(0.00)} & 0.89(0.67) \\ 
ED  & \textbf{99.97(0.24)} & 0.00(0.00) & \textbf{98.71(1.80)} & 0.00(0.00) & 0.00(0.00) & 0.00(0.00) & 21.52(7.10) & 0.00(0.00) \\ 
SEQ1  & \textbf{99.98(0.22)} & 0.00(0.00) & \textbf{100.00(0.00)} & 0.00(0.02) & 6.35(4.97) & 0.00(0.00) & \textbf{100.00(0.00)} & 0.00(0.00) \\ 
  SEQ2  & \textbf{99.98(0.22)} & 0.68(0.52) & \textbf{100.00(0.00)} & 0.51(0.46) & 24.55(11.66) & 0.00(0.00) & 81.52(6.46) & 0.48(0.43) \\ 
  SEQ3  & \textbf{99.98(0.22)} & 0.00(0.00) & \textbf{100.00(0.00)} & 0.00(0.02) & 3.46(3.21) & 0.00(0.00) & \textbf{100.00(0.00)} & 0.00(0.00) \\ 
\end{tabular}}
\setlength\tabcolsep{15pt} 
{\renewcommand{\arraystretch}{.7}
\scriptsize
\begin{tabular}{@{}lllllll@{}}
  \toprule
 \multirow{2}{*}{Method} & \multicolumn{2}{c}{Model 6} & \multicolumn{2}{c}{Model 7} & \multicolumn{2}{c}{Model 8} \\
 \cmidrule{2-3} \cmidrule{4-5} \cmidrule{6-7}
 & TPR & FPR & TPR & FPR & TPR & FPR \\ 
  \midrule
\textit{FST} & \textbf{79.53(9.87)} & 4.58(1.32) & 41.90(20.48) & 6.75(2.03) & \textbf{96.45(3.47)} & 5.08(1.28) \\ 
\textit{FSTMG}  & 0.98(1.50) & 1.74(1.04) & 1.67(1.97) & 1.68(0.93) & 30.07(6.58) & 0.81(0.63) \\ 
\textit{FSTSH} & \textbf{77.73(9.75)} & 2.67(0.89) & 4.06(3.04) & 5.16(1.82) & 68.58(6.13) & 3.95(1.15) \\ 
\textit{FSTAM} & 2.98(2.84) & 1.18(0.80) & 39.66(21.37) & 0.00(0.00) & 9.31(4.30) & 1.20(0.80) \\ 
\textit{SF} & \textbf{79.25(9.73)} & 4.24(1.21) & 34.30(15.56) & 6.77(2.09) & \textbf{95.18(4.01)} & 4.65(1.18) \\ 
\textit{SF25} & \textbf{79.05(10.23)} & 4.20(1.24) & 33.84(16.75) & 6.83(2.04) & \textbf{95.04(4.18)} & 4.63(1.23) \\ 
\textit{MUOD} & 39.41(20.18) & 11.68(4.77) & 92.22(12.04) & 18.09(6.89) & 56.48(15.12) & 3.75(2.44) \\ 
OGMBD & \textbf{98.72(2.08)} & 0.75(0.58) & 3.33(5.64) & 0.00(0.00) & 80.19(6.37) & 2.15(0.96) \\ 
MSPLT & \textbf{100.00(0.00)} & 2.44(1.24) & 52.31(17.11) & 0.02(0.10) & \textbf{100.00(0.10)} & 2.43(1.24) \\ 
TVD & 55.78(19.38) & 0.00(0.00) & 20.12(12.48) & 0.00(0.00) & \textbf{98.59(2.12)} & 0.00(0.02) \\ 
FOM & 0.00(0.10) & 0.02(0.09) & 0.01(0.17) & 0.00(0.00) & 33.43(8.57) & 0.02(0.09) \\ 
FAO & 0.00(0.00) & 0.00(0.04) & 0.03(0.30) & 0.00(0.00) & 28.36(7.21) & 0.00(0.04) \\ 
FOM2 & 39.80(16.33) & 1.09(0.66) & 5.30(5.49) & 0.06(0.15) & \textbf{97.17(3.27)} & 1.12(0.70) \\ 
FAO2 & 28.24(15.09) & 0.78(0.62) & 1.53(3.10) & 0.05(0.15) & \textbf{92.58(4.88)} & 0.89(0.66) \\ 
ED & 0.00(0.00) & 0.00(0.00) & 0.00(0.00) & 0.00(0.00) & 55.08(7.47) & 0.00(0.00) \\ 
SEQ1 & 0.10(0.46) & 0.00(0.00) & 0.00(0.00) & 0.00(0.00) & 75.32(6.27) & 0.00(0.00) \\ 
SEQ2 & 5.58(4.28) & 0.48(0.43) & 1.61(1.96) & 0.01(0.05) & 74.28(6.67) & 0.55(0.48) \\ 
SEQ3 & 0.00(0.00) & 0.00(0.00) & 0.00(0.00) & 0.00(0.00) & 74.78(6.30) & 0.00(0.00) \\ 
 \bottomrule
\end{tabular}}
\end{table*}

\begin{table*}[h!]

\centering
\caption{\label{tab:sim-results-300-50-0.2}Mean and Standard Deviation (in parentheses) of the True Positive Rates (TPR) and False Positive Rate (FPR) over eight simulation models with 500 repetitions for each possible case. Each simulation is done with $n=300$ and $d=50$ and $\alpha = 0.2$. Comparatively high TPRs are in bold. Proposed methods in italics.}
{\renewcommand{\arraystretch}{.7}
\scriptsize
\begin{tabular}{@{}lllllllll@{}} 
\toprule
\multirow{2}{*}{Method} & \multicolumn{2}{c}{Model 2}  & \multicolumn{2}{c}{Model 3} & \multicolumn{2}{c}{Model 4} & \multicolumn{2}{c}{Model 5} \\
\cmidrule{2-3} \cmidrule{4-5} \cmidrule{6-7} \cmidrule{8-9}	

& TPR & FPR & TPR & FPR & TPR & FPR & TPR & FPR \\ 
 \midrule
\textit{FST} & \textbf{99.96(0.26)} & 8.84(1.65) & 73.15(13.34) & 2.91(1.14) & \textbf{99.69(0.74)} & 0.44(0.46) & 75.13(8.18) & 2.35(0.91) \\ 
\textit{FSTMG}  & \textbf{99.92(0.36)} & 0.00(0.03) & 3.72(2.59) & 1.40(0.86) & 22.86(6.32) & 0.21(0.31) & 13.92(4.77) & 0.68(0.58) \\ 
\textit{FSTSH} & 7.89(3.46) & 7.98(1.59) & 69.49(13.42) & 1.16(0.63) & \textbf{99.67(0.76)} & 0.19(0.30) & 62.95(7.98) & 1.54(0.71) \\ 
\textit{FSTAM} & 1.62(1.70) & 1.70(0.94) & 5.41(3.05) & 1.14(0.79) & 13.12(6.18) & 0.06(0.15) & 19.26(5.40) & 0.52(0.52) \\ 
\textit{SF} & \textbf{99.74(1.14)} & 8.56(1.60) & 71.60(13.23) & 2.55(1.07) & 86.94(8.61) & 0.20(0.29) & 69.35(8.24) & 2.15(0.88) \\ 
\textit{SF25} & \textbf{99.46(1.95)} & 8.54(1.67) & 71.24(12.98) & 2.54(1.10) & 84.93(13.91) & 0.20(0.31) & 69.18(8.11) & 2.14(0.88) \\ 
\textit{MUOD} & \textbf{100.00(0.00)} & 8.31(3.92) & 40.59(21.34) & 10.16(4.67) & 74.23(18.99) & 3.42(4.24) & 37.97(11.39) & 3.24(2.43) \\ 
OGMBD  & \textbf{99.95(0.33)} & 4.52(1.48) & 22.58(8.24) & 2.62(1.08) & 7.21(5.31) & 0.07(0.19) & 75.17(8.95) & 0.30(0.36) \\ 
MSPLT   & \textbf{99.90(0.44)} & 2.01(1.16) & \textbf{100.00(0.00)} & 2.06(1.16) & \textbf{99.34(1.12)} & 0.97(0.73) & \textbf{99.98(0.17)} & 2.05(1.04) \\ 
TVD  & \textbf{99.95(0.33)} & 0.00(0.03) & \textbf{100.00(0.00)} & 0.00(0.00) & 0.75(1.26) & 0.00(0.00) & \textbf{100.00(0.00)} & 0.00(0.00) \\ 
FOM  & \textbf{99.99(0.13)} & 0.00(0.00) & 4.63(7.73) & 0.00(0.04) & 0.51(1.06) & 0.00(0.03) & 1.39(1.87) & 0.00(0.02) \\ 
FAO & \textbf{99.88(0.67)} & 0.00(0.00) & 1.30(3.08) & 0.00(0.00) & 0.05(0.28) & 0.05(0.16) & 0.55(1.17) & 0.00(0.00) \\ 
FOM2  & \textbf{100.00(0.00)} & 0.14(0.26) & \textbf{100.00(0.00)} & 0.72(0.60) & 12.28(4.90) & 0.50(0.51) & \textbf{100.00(0.00)} & 0.53(0.51) \\ 
FAO2 & \textbf{100.00(0.00)} & 0.26(0.35) & \textbf{100.00(0.00)} & 0.44(0.46) & 1.39(1.72) & 2.25(1.20) & \textbf{100.00(0.00)} & 0.39(0.46) \\ 
ED  & \textbf{99.93(0.53)} & 0.00(0.00) & \textbf{97.60(2.48)} & 0.00(0.00) & 0.00(0.00) & 0.00(0.00) & 17.78(6.05) & 0.00(0.00) \\ 
SEQ1  & \textbf{99.96(0.27)} & 0.00(0.00) & \textbf{100.00(0.00)} & 0.00(0.00) & 4.12(3.56) & 0.00(0.00) & \textbf{100.00(0.00)} & 0.00(0.00) \\ 
  SEQ2  & \textbf{99.96(0.27)} & 0.67(0.52) & \textbf{100.00(0.00)} & 0.44(0.41) & 16.55(9.34) & 0.00(0.00) & 78.28(6.15) & 0.46(0.41) \\ 
  SEQ3 & \textbf{99.96(0.27)} & 0.00(0.00) & \textbf{100.00(0.00)} & 0.00(0.00) & 2.40(2.58) & 0.00(0.00) & \textbf{100.00(0.00)} & 0.00(0.00) \\ 
\end{tabular}}
\setlength\tabcolsep{15.5pt} 
{\renewcommand{\arraystretch}{.7}
\scriptsize

\begin{tabular}{@{}lllllll@{}}
  \toprule
  \multirow{2}{*}{Method} & \multicolumn{2}{c}{Model 6} & \multicolumn{2}{c}{Model 7} & \multicolumn{2}{c}{Model 8}\\
  \cmidrule{2-3} \cmidrule{4-5} \cmidrule{6-7}
& TPR & FPR & TPR & FPR & TPR & FPR \\ 
  \hline
\textit{FST} & 55.81(9.97) & 3.30(1.12) & 9.32(6.13) & 7.18(2.03) & \textbf{91.60(5.80)} & 3.66(1.13) \\ 
\textit{FSTMG} & 0.91(1.21) & 1.79(0.91) & 1.76(1.75) & 1.70(0.92) & 28.88(5.48) & 0.62(0.55) \\ 
\textit{FSTSH}  & 54.20(9.69) & 1.29(0.70) & 4.32(2.62) & 5.55(1.84) & 63.16(6.13) & 2.70(1.01) \\ 
\textit{FSTAM} & 2.62(2.30) & 1.05(0.74) & 4.20(6.36) & 0.00(0.00) & 9.20(3.87) & 0.96(0.70) \\ 
\textit{SF}& 54.30(10.28) & 3.07(1.08) & 7.99(5.07) & 7.16(2.03) & \textbf{89.58(6.17)} & 3.21(1.09) \\ 
\textit{SF25} & 54.55(10.22) & 3.05(1.06) & 9.08(6.27) & 7.20(2.09) & \textbf{89.59(6.21)} & 3.23(1.08) \\ 
\textit{MUOD} & 33.77(19.12) & 12.08(5.16) & 85.97(14.26) & 16.82(7.26) & 51.12(13.75) & 3.53(2.97) \\ 
OGMBD & \textbf{84.93(14.27)} & 0.13(0.23) & 0.13(0.65) & 0.00(0.00) & 76.86(6.27) & 1.51(0.88) \\ 
MSPLT & 100.00(0.00) & 1.96(1.16) & 38.82(16.58) & 0.02(0.10) & \textbf{99.96(0.24)} & 2.03(1.08) \\ 
TVD & 12.77(12.73) & 0.00(0.00) & 2.89(5.08) & 0.00(0.00) & \textbf{95.57(4.18)} & 0.00(0.02) \\ 
FOM & 0.00(0.00) & 0.01(0.06) & 0.00(0.00) & 0.00(0.00) & 27.57(6.55) & 0.00(0.00) \\ 
FAO & 0.00(0.00) & 0.00(0.03) & 0.00(0.00) & 0.00(0.00) & 25.33(5.84) & 0.00(0.00) \\ 
FOM2 & 13.39(9.51) & 0.50(0.48) & 0.33(0.80) & 0.01(0.08) & \textbf{92.76(5.06)} & 0.56(0.53) \\ 
FAO2 & 8.26(6.89) & 0.33(0.40) & 0.14(0.85) & 0.03(0.13) & 86.48(5.47) & 0.43(0.47) \\ 
ED & 0.00(0.00) & 0.00(0.00) & 0.00(0.00) & 0.00(0.00) & 53.28(6.71) & 0.00(0.00) \\ 
SEQ1 & 0.10(0.48) & 0.00(0.00) & 0.00(0.00) & 0.00(0.00) & 75.53(5.80) & 0.00(0.00) \\ 
SEQ2 & 4.49(3.51) & 0.45(0.41) & 1.65(1.61) & 0.00(0.03) & 73.00(6.24) & 0.46(0.42) \\ 
SEQ3 & 0.00(0.00) & 0.00(0.00) & 0.00(0.00) & 0.00(0.00) & 75.14(5.80) & 0.00(0.00) \\ 
 \bottomrule
\end{tabular}}
\end{table*}

\begin{table*}[h!]
\caption{\label{tab:sim-results-100-25}Mean and Standard Deviation (in parentheses) of the True Positive Rate (TPR) and the False Positive Rate (FPR) over eight simulation models with 500 repetitions for each possible case with sample size $n=100$, evaluation points $d=25$, and contamination rate $\alpha = 0.1$. Comparatively high TPRs are marked in bold.}
\centering

{\renewcommand{\arraystretch}{.7}
  \footnotesize
\begin{tabular}{@{}lccccccc@{}}  \toprule
  \multirow{2}{*}{Method} & Model1  & \multicolumn{2}{c}{Model 2} & \multicolumn{2}{c}{Model 3} & \multicolumn{2}{c}{Model 4} \\
  \cmidrule{2-2} \cmidrule{3-4} \cmidrule{5-6} \cmidrule{7-8}
& FPR & TPR & FPR & TPR & FPR & TPR & FPR \\ 
  \midrule
\textit{FST} & 10.07(2.87) & \textbf{100.00(0.00)} & 8.99(2.81) & \textbf{99.04(3.51)} & 6.16(2.37) & \textbf{99.98(0.45)} & 3.15(1.85) \\ 
\textit{FSTMG} & 1.94(1.62) & \textbf{100.00(0.00)} & 0.45(0.78) & 4.66(6.59) & 1.64(1.49) & 39.70(16.82) & 0.69(0.93) \\ 
\textit{FSTSH} & 7.93(2.48) & 7.70(8.73) & 7.86(2.63) & \textbf{98.08(4.85)} & 4.34(1.95) & \textbf{99.98(0.45)} & 2.17(1.51) \\ 
\textit{FSTAM} &  1.97(1.73) & 2.04(4.63) & 1.83(1.62) & 6.70(8.33) & 1.43(1.43) & 50.26(19.82) & 0.50(0.76) \\ 
\textit{SF} & 9.76(2.83) & \textbf{99.96(0.63)} & 8.62(2.81) & \textbf{98.50(4.43)} & 5.77(2.24) & \textbf{99.90(1.00)} & 2.68(1.69) \\ 
\textit{SF25} & 9.72(2.76) & \textbf{99.58(3.85)} & 8.77(2.87) & \textbf{98.30(4.92)} & 5.62(2.22) & \textbf{99.34(3.37)} & 2.66(1.72) \\ 
\textit{MUOD} & 29.10(12.38) & 99.64(3.89) & 18.92(10.08) & 85.70(19.89) & 25.50(14.10) & \textbf{99.00(4.36)} & 11.16(6.35) \\ 
OGMBD & 5.25(2.31) & \textbf{100.00(0.00)} & 4.66(2.35) & 51.96(17.84) & 3.85(2.13) & \textbf{94.68(7.89)} & 1.33(1.18) \\ 
MSPLOT & 2.97(2.37) & \textbf{99.28(2.88)} & 2.12(1.86) & \textbf{100.00(0.00)} & 2.11(1.98) & \textbf{99.82(1.33)} & 0.90(1.16) \\ 
TVD & 0.04(0.22) & \textbf{100.00(0.00)} & 0.02(0.16) & \textbf{100.00(0.00)} & 0.02(0.16) & 15.66(14.62) & 0.00(0.00) \\ 
FOM & 0.69(1.01) & \textbf{100.00(0.00)} & 0.09(0.39) & 40.30(26.34) & 0.14(0.46) & 49.16(26.10) & 0.12(0.43) \\ 
FAO & 0.32(0.68) & \textbf{100.00(0.00)} & 0.06(0.32) & 22.58(20.10) & 0.04(0.22) & 12.20(19.55) & 0.03(0.20) \\ 
FOM2 & 2.06(1.62) & \textbf{100.00(0.00)} & 0.45(0.76) & \textbf{100.00(0.00)} & 0.92(1.10) & 66.64(20.43) & 0.36(0.71) \\ 
FAO2 & 1.68(1.59) & \textbf{100.00(0.00)} & 0.64(0.99) & \textbf{100.00(0.00) }& 0.66(0.99) & 27.54(24.54) & 0.29(0.63) \\ 
ED & 0.01(0.11) & \textbf{100.00(0.00)} & 0.01(0.12) & \textbf{99.26(3.04)} & 0.02(0.13) & 5.22(8.74) & 0.00(0.00) \\ 
SEQ1 & 0.03(0.19) & \textbf{100.00(0.00)} & 0.02(0.16) & \textbf{100.00(0.00)} & 0.03(0.17) & 38.96(20.61) & 0.00(0.00) \\ 
SEQ2 & 1.27(1.17) & \textbf{100.00(0.00)} & 1.24(1.18) & \textbf{100.00(0.00)} & 1.08(1.09) & 69.76(18.96) & 0.00(0.00) \\ 
SEQ3 & 0.03(0.18) & \textbf{100.00(0.00)} & 0.02(0.15) & \textbf{100.00(0.00)} & 0.02(0.16) & 18.76(14.49) & 0.00(0.00) \\ 

\end{tabular}}

\setlength\tabcolsep{2.5pt} 
{\renewcommand{\arraystretch}{.7}
  \footnotesize
\begin{tabular}{@{}lcccccccc@{}}  \toprule
\multirow{2}{*}{Method} & \multicolumn{2}{c}{Model 5} & \multicolumn{2}{c}{Model 6} & \multicolumn{2}{c}{Model 7} & \multicolumn{2}{c}{Model 8} \\
  \cmidrule{2-3} \cmidrule{4-5} \cmidrule{6-7} \cmidrule{8-9}
& TPR & FPR & TPR & FPR & TPR & FPR & TPR & FPR\\ 

  \midrule
\textit{FST} & \textbf{94.18(8.37)} & 5.93(2.26) & \textbf{91.92(11.89)} & 6.37(2.32) & \textbf{73.42(26.24)} & 6.96(3.17) & \textbf{97.56(5.52)} & 6.80(2.38) \\ 
\textit{FSTMG} & 17.14(12.05) & 1.30(1.31) & 0.88(2.91) & 1.77(1.63) & 1.62(4.10) & 1.83(1.67) & 30.86(14.54) & 1.18(1.30) \\ 
\textit{FSTSH} & 83.54(13.01) & 4.47(1.90) & \textbf{90.12(12.44)} & 4.38(1.82) & 4.48(6.63) & 5.20(2.87) & 70.62(14.58) & 5.24(2.14) \\ 
\textit{FSTAM} & 23.28(13.22) & 1.12(1.23) & 3.96(6.39) & 1.50(1.38) & \textbf{72.56(26.80)} & 0.03(0.21) & 9.70(9.31) & 1.42(1.38) \\ 
\textit{SF} & \textbf{91.94(9.73)} & 5.43(2.13) & \textbf{92.14(11.36)} & 5.91(2.21) & 62.74(26.88) & 7.00(3.00) & \textbf{97.30(5.81)} & 6.30(2.25) \\ 
\textit{SF25} & \textbf{91.80(9.93)} & 5.38(2.08) & \textbf{90.46(12.44)} & 5.87(2.28) & 60.36(27.69) & 6.97(3.01) & \textbf{96.86(6.39)} & 6.24(2.35) \\ 
\textit{MUOD} & 80.92(17.19) & 16.44(10.97) & 80.26(21.09) & 26.72(12.57) & 98.32(6.76) & 33.58(12.18) & 88.32(14.55) & 15.26(11.95) \\ 
OGMBD & \textbf{96.68(5.89)} & 2.18(1.62) & \textbf{99.88(1.09)} & 2.05(1.57) & 22.72(25.78) & 0.00(0.07) & 86.22(11.18) & 2.85(1.86) \\ 
MSPLT & \textbf{99.42(2.42)} & 2.04(1.75) & \textbf{100.00(0.00)} & 2.16(1.95) & 48.42(27.14) & 0.03(0.20) & \textbf{99.72(1.65)} & 2.09(1.86) \\ 
TVD & \textbf{100.00(0.00)} & 0.03(0.17) & 86.00(20.49) & 0.02(0.15) & 27.38(24.90) & 0.00(0.05) & \textbf{97.98(5.23)} & 0.02(0.16) \\ 
FOM & 13.12(14.69) & 0.12(0.39) & 0.16(1.26) & 0.16(0.45) & 2.78(8.55) & 0.01(0.09) & 39.06(17.63) & 0.15(0.47) \\ 
FAO & 8.04(11.38) & 0.06(0.27) & 0.14(1.34) & 0.06(0.24) & 2.58(7.30) & 0.01(0.11) & 33.42(16.24) & 0.06(0.27) \\ 
FOM2 & \textbf{100.00(0.00)} & 0.94(1.07) & \textbf{89.54(16.99)} & 0.88(1.06) & 17.40(19.86) & 0.16(0.44) & \textbf{99.46(2.51)} & 0.80(0.98) \\ 
FAO2 & \textbf{100.00(0.00)} & 0.63(0.93) & 70.16(27.97) & 0.65(0.92) & 12.90(17.81) & 0.15(0.47) & \textbf{95.94(7.20)} & 0.62(0.91) \\ 
ED & 31.16(16.07) & 0.00(0.05) & 0.14(1.34) & 0.02(0.16) & 0.00(0.00) & 0.00(0.00) & 57.62(15.55) & 0.01(0.09) \\ 
SEQ1 & \textbf{99.86(1.18)} & 0.01(0.10) & 7.00(9.80) & 0.02(0.16) & 0.04(0.63) & 0.00(0.00) & 79.36(12.13) & 0.01(0.11) \\ 
SEQ2 & 87.98(10.79) & 1.05(1.05) & 28.34(17.62) & 1.14(1.10) & 3.24(5.51) & 0.06(0.26) & 82.66(11.76) & 1.07(1.10) \\ 
SEQ3 & \textbf{99.94(0.77)} & 0.01(0.09) & 0.20(1.40) & 0.02(0.16) & 0.04(0.63) & 0.00(0.00) & 74.98(13.09) & 0.01(0.11) \\ 
   \bottomrule
\end{tabular}}
\end{table*}

\begin{table*}[h!]
\centering
\caption{\label{tab:sim-results-correlation}\textit{Mean and Standard Deviation (in parentheses) of the TPR and FPR over three models with 500 repetitions for each possible case with $n=300$, $d=50$,  $\alpha = 0.1$. Comparatively high TPRs are marked in bold.}}
\setlength\tabcolsep{3.0pt} 
{\renewcommand{\arraystretch}{.7}
  \footnotesize
\begin{tabular}{@{}lcccccc@{}}  \toprule
\multirow{2}{*}{Method} & \multicolumn{2}{c}{Model 3} & \multicolumn{2}{c}{Model 4} & \multicolumn{2}{c}{Model 6} \\
  \cmidrule{2-3} \cmidrule{4-5} \cmidrule{6-7} 
& TPR & FPR & TPR & FPR & TPR & FPR \\ 
\midrule
FSTSH\_PEARSON & \textbf{99.03(2.17)} & 4.36(1.16) & \textbf{99.99(0.15)} & 2.12(0.89) & \textbf{89.80(6.86)} & 4.38(1.12) \\ 
FSTSH\_KENDALL & 8.73(5.36) & 4.48(1.26) & \textbf{99.89(0.59)} & 0.64(0.52) & 76.59(11.80) & 2.66(0.97) \\ 
FSTSH\_SPEARMAN & 29.67(9.97) & 6.67(1.41) & \textbf{99.97(0.33)} & 1.79(0.82) & \textbf{87.25(8.73)} & 4.87(1.23) \\ 
FSTSH\_COSINE & 71.09(11.17) & 6.58(1.17) & \textbf{99.95(0.42)} & 2.30(0.97) & 54.07(11.80) & 6.78(1.27) \\ 
   \bottomrule
\end{tabular}}
\end{table*}

\begin{table*}[h!]
\centering
\caption{\label{tab:snr1}Mean and Standard Deviation (in parentheses) of the TPR and FPR over four models with 500 repetitions for each possible case with $n=300$, $d=50$,  $\alpha = 0.1$ and $\nu \in \{0.25, 0.5\}$. Comparatively high TPRs are marked in bold.}
\setlength\tabcolsep{3.0pt} 
{\renewcommand{\arraystretch}{.7}
  \footnotesize
\begin{tabular}{@{}lcccccccc@{}}  \toprule
\multirow{2}{*}{Method} & \multicolumn{2}{c}{Model 2} & \multicolumn{2}{c}{Model 3} & \multicolumn{2}{c}{Model 4} & \multicolumn{2}{c}{Model 6} \\
  \cmidrule{2-3} \cmidrule{4-5} \cmidrule{6-7} \cmidrule{8-9}
& TPR & FPR & TPR & FPR & TPR & FPR & TPR & FPR\\ 
\midrule
   \multicolumn{9}{c}{$\nu = 0.25$}\\
\hline
\textit{FST} & \textbf{100.00(0.00)} & 6.92(1.59) & \textbf{100.00(0.00)} & 4.29(1.21) & \textbf{100.00(0.00)} & 3.48(1.16) & \textbf{100.00(0.00)} & 4.01(1.21) \\ 
\textit{FSTMG} & \textbf{100.00(0.00)} & 0.39(0.41) & 12.06(6.40) & 1.22(0.75) & 38.07(10.76) & 0.53(0.46) & 11.65(6.27) & 1.09(0.73) \\ 
\textit{FSTSH} & 5.88(4.44) & 5.49(1.45) & \textbf{100.00(0.00)} & 2.63(0.97) & \textbf{100.00(0.00)} & 2.68(1.01) & \textbf{100.00(0.00)} & 2.58(0.96) \\ 
\textit{FSTAM} & 1.70(2.44) & 1.71(0.97) & 21.58(7.67) & 0.99(0.70) & \textbf{95.71(3.99)} & 0.38(0.39) & 30.75(8.73) & 0.73(0.58) \\ 
\textit{SF} & \textbf{100.00(0.00)} & 6.92(1.58) & \textbf{100.00(0.00)} & 4.08(1.17) & \textbf{100.00(0.00)} & 2.73(1.09) & \textbf{100.00(0.00)} & 3.85(1.18) \\ 
\textit{SF25} & \textbf{100.00(0.00)} & 6.93(1.65) & \textbf{100.00(0.00)} & 4.10(1.21) & \textbf{100.00(0.00)} & 2.65(1.07) & \textbf{100.00(0.00)} & 3.81(1.17) \\ 
\textit{MUOD} & \textbf{100.00(0.00)} & 10.11(4.09) & \textbf{99.11(6.77)} & 6.68(3.35) & \textbf{99.98(0.45)} & 5.52(2.93) & 94.50(14.33) & 6.76(3.55) \\ 
OGMBD & \textbf{100.00(0.00)} & 4.76(1.42) & 38.89(11.78) & 3.54(1.23) & 100.00(0.00) & 2.08(0.87) & \textbf{100.00(0.00)} & 1.71(0.86) \\ 
MSPLT & \textbf{100.00(0.00)} & 2.95(1.32) & \textbf{100.00(0.00)} & 2.99(1.31) & 100.00(0.00) & 2.91(1.38) & \textbf{100.00(0.00)} & 2.81(1.33) \\ 
TVD & \textbf{100.00(0.00)} & 0.00(0.02) & \textbf{100.00(0.00)} & 0.00(0.00) & \textbf{99.26(1.89)} & 0.00(0.00) & \textbf{100.00(0.00)} & 0.00(0.03) \\ 
FOM & \textbf{100.00(0.00)} & 0.06(0.16) & \textbf{99.97(0.33)} & 0.11(0.22) & 7.19(10.03) & 0.05(0.15) & 44.57(31.21) & 0.08(0.20) \\ 
FAO & \textbf{100.00(0.00)} & 0.02(0.09) & \textbf{99.55(1.45)} & 0.03(0.10) & 0.07(0.51) & 0.00(0.04) & 19.87(19.94) & 0.03(0.12) \\ 
FOM2 & \textbf{100.00(0.00)} & 1.26(0.71) & \textbf{100.00(0.00)} & 2.06(0.99) & \textbf{99.49(1.32)} & 1.75(0.86) & \textbf{100.00(0.00)} & 1.87(0.86) \\ 
FAO2 & \textbf{100.00(0.00)} & 1.41(0.79) & \textbf{100.00(0.00)} & 1.61(0.91) & 39.20(23.27) & 1.26(0.81) & \textbf{100.00(0.00)} & 1.53(0.81) \\ 
ED & \textbf{100.00(0.00)} & 0.00(0.00) & \textbf{100.00(0.00)} & 0.00(0.00) & 11.15(9.54) & 0.00(0.00) & 26.19(11.04) & 0.00(0.02) \\ 
SEQ1 & \textbf{100.00(0.00)} & 0.00(0.00) & \textbf{100.00(0.00)} & 0.00(0.00) & \textbf{98.80(2.17)} & 0.00(0.00) & \textbf{99.99(0.15)} & 0.00(0.02) \\ 
SEQ2 & \textbf{100.00(0.00)} & 0.00(0.02) & \textbf{100.00(0.00)} & 0.01(0.04) & \textbf{99.95(0.39)} & 0.00(0.02) & \textbf{100.00(0.00)} & 0.01(0.04) \\ 
SEQ3 & \textbf{100.00(0.00)} & 0.00(0.00) & \textbf{100.00(0.00)} & 0.00(0.00) & 31.67(13.54) & 0.00(0.00) & 26.29(11.12) & 0.00(0.02) \\ 
   \midrule
   \multicolumn{9}{c}{$\nu = 0.5$}\\
   \midrule
\textit{FST} & \textbf{100.00(0.00)} & 7.62(1.54) & \textbf{100.00(0.00)} & 5.03(1.27) & \textbf{99.99(0.21)} & 3.64(1.17) & \textbf{100.00(0.00)} & 4.82(1.23) \\ 
\textit{FSTMG} & \textbf{100.00(0.00)} & 0.38(0.38) & 6.62(4.57) & 1.40(0.82) & 17.77(8.22) & 0.82(0.58) & 6.08(4.48) & 1.31(0.81) \\ 
\textit{FSTSH} & 6.54(4.52) & 6.44(1.41) & \textbf{100.00(0.00)} & 3.26(1.01) & \textbf{99.99(0.21)} & 2.58(1.01) & \textbf{100.00(0.00)} & 3.23(0.97) \\ 
\textit{FSTAM} & 1.63(2.38) & 1.73(0.97) & 11.67(6.13) & 1.20(0.82) & 60.70(11.91) & 0.36(0.38) & 15.65(6.69) & 1.01(0.70) \\ 
\textit{SF} & \textbf{100.00(0.00)} & 7.58(1.60) & \textbf{100.00(0.00)} & 4.73(1.24) & \textbf{99.96(0.36)} & 3.11(1.11) & \textbf{100.00(0.00)} & 4.58(1.19) \\ 
\textit{SF25} & \textbf{100.00(0.00)} & 7.55(1.54) & \textbf{100.00(0.00)} & 4.71(1.19) & \textbf{99.94(0.44)} & 3.15(1.14) & \textbf{100.00(0.00)} & 4.59(1.21) \\ 
\textit{MUOD} & \textbf{100.00(0.00)} & 9.50(3.90) & 87.49(20.59) & 9.51(4.31) & \textbf{96.41(8.80)} & 6.15(3.63) & 69.81(22.80) & 8.45(4.16) \\ 
OGMBD & \textbf{100.00(0.00)} & 4.84(1.45) & 38.75(11.62) & 3.48(1.21) & \textbf{98.57(2.28)} & 2.02(0.88) & \textbf{100.00(0.00)} & 1.78(0.82) \\ 
MSPLT & \textbf{100.00(0.00)} & 2.97(1.39) & \textbf{100.00(0.00)} & 2.87(1.35) & \textbf{99.97(0.30)} & 2.83(1.32) & \textbf{100.00(0.00)} & 2.94(1.33) \\ 
TVD & \textbf{100.00(0.00)} & 0.00(0.02) & \textbf{100.00(0.00)} & 0.00(0.00) & 38.87(18.45) & 0.00(0.00) & \textbf{100.00(0.00)} & 0.00(0.02) \\ 
FOM & \textbf{100.00(0.00)} & 0.07(0.17) & \textbf{93.29(6.72)} & 0.09(0.19) & 0.65(1.69) & 0.06(0.15) & 2.85(4.28) & 0.06(0.17) \\ 
FAO & \textbf{100.00(0.00)} & 0.02(0.11) & 82.73(12.58) & 0.03(0.10) & 0.11(0.64) & 0.00(0.04) & 2.01(3.48) & 0.01(0.09) \\ 
FOM2 & \textbf{100.00(0.00)} & 1.30(0.72) & \textbf{100.00(0.00)} & 1.95(0.88) & 66.22(13.00) & 1.80(0.87) & \textbf{100.00(0.00)} & 1.89(0.87) \\ 
FAO2 & \textbf{100.00(0.00)} & 1.42(0.82) & \textbf{100.00(0.00)} & 1.49(0.74) & 17.43(13.20) & 1.50(0.85) & \textbf{100.00(0.00)} & 1.53(0.84) \\ 
ED & \textbf{100.00(0.00)} & 0.00(0.00) & \textbf{100.00(0.00)} & 0.00(0.00) & 0.79(1.77) & 0.00(0.00) & 2.22(2.82) & 0.00(0.00) \\ 
SEQ1 & \textbf{100.00(0.00)} & 0.00(0.00) & \textbf{100.00(0.00)} & 0.00(0.00) & 41.70(14.60) & 0.00(0.00) & 49.93(13.31) & 0.00(0.00) \\ 
SEQ2 & \textbf{100.00(0.00)} & 0.09(0.18) & \textbf{100.00(0.00)} & 0.07(0.17) & 80.37(10.31) & 0.00(0.03) & 72.23(10.45) & 0.07(0.16) \\ 
SEQ3 & \textbf{100.00(0.00)} & 0.00(0.00) & \textbf{100.00(0.00)} & 0.00(0.00) & 2.81(3.44) & 0.00(0.00) & 2.39(2.93) & 0.00(0.00) \\ 
\hline

\end{tabular}}
\end{table*}

\begin{table*}[h!]
\centering
\caption{\label{tab:snr2}Mean and Standard Deviation (in parentheses) of the TPR and FPR over four models with 500 repetitions for each possible case with $n=300$, $d=50$,  $\alpha = 0.1$ and $\nu \in \{1.5, 5\}$. Comparatively high TPRs are marked in bold.}
\setlength\tabcolsep{3.0pt} 
{\renewcommand{\arraystretch}{.7}
  \footnotesize
\begin{tabular}{@{}lcccccccc@{}}  \toprule
\multirow{2}{*}{Method} & \multicolumn{2}{c}{Model 2} & \multicolumn{2}{c}{Model 3} & \multicolumn{2}{c}{Model 4} & \multicolumn{2}{c}{Model 6} \\
  \cmidrule{2-3} \cmidrule{4-5} \cmidrule{6-7} \cmidrule{8-9}
& TPR & FPR & TPR & FPR & TPR & FPR & TPR & FPR\\ 
\midrule

\multicolumn{9}{c}{$\nu = 1.5$}\\
  \midrule
\textit{FST} & \textbf{99.93(0.49)} & 10.06(1.55) & 80.99(9.75) & 6.86(1.35) & \textbf{90.73(6.01)} & 4.43(1.39) & 65.28(11.38) & 7.10(1.37) \\ 
\textit{FSTMG} & \textbf{99.89(0.60)} & 0.35(0.38) & 3.23(3.33) & 1.53(0.86) & 5.97(4.60) & 1.38(0.87) & 3.09(3.16) & 1.59(0.85) \\ 
\textit{FSTSH} & 8.93(5.00) & 9.03(1.45) & 78.13(9.77) & 5.13(1.21) & \textbf{90.65(6.08)} & 2.52(0.97) & 61.44(10.96) & 5.32(1.18) \\ 
\textit{FSTAM} & 1.71(2.38) & 1.69(0.91) & 4.77(3.94) & 1.43(0.84) & 18.06(8.08) & 0.81(0.61) & 5.66(4.45) & 1.37(0.79) \\ 
\textit{SF} & \textbf{99.94(0.44)} & 9.46(1.51) & 79.93(9.68) & 6.30(1.31) & \textbf{87.32(6.69)} & 3.64(1.19) & 63.53(11.06) & 6.47(1.28) \\ 
 \textit{SF25} & \textbf{99.91(0.57)} & 9.49(1.49) & 79.45(9.82) & 6.28(1.38) & \textbf{87.04(8.02)} & 3.72(1.24) & 63.49(10.80) & 6.48(1.26) \\ 
 \textit{MUOD} & \textbf{97.78(8.96)} & 8.65(4.38) & 34.67(17.21) & 11.26(4.68) & 68.05(15.45) & 8.30(4.04) & 31.61(14.09) & 11.05(4.57) \\ 
 OGMBD & \textbf{98.25(2.51)} & 4.72(1.39) & 32.25(10.39) & 3.42(1.13) & 67.01(9.08) & 2.24(0.95) & \textbf{88.95(7.15)} & 1.93(0.85) \\ 
MSPLT & \textbf{94.21(5.34)} & 2.92(1.34) & \textbf{100.00(0.00)} & 2.87(1.34) & \textbf{80.71(8.70)} & 2.92(1.34) & \textbf{99.12(2.00)} & 2.79(1.27) \\ 
TVD & \textbf{98.24(2.56)} & 0.00(0.02) & \textbf{100.00(0.00)} & 0.00(0.00) & 0.44(1.26) & 0.00(0.02) & 54.42(15.48) & 0.00(0.00) \\ 
FOM & \textbf{99.99(0.15)} & 0.06(0.16) & 13.05(12.26) & 0.11(0.23) & 0.58(1.52) & 0.17(0.26) & 0.43(1.22) & 0.12(0.22) \\  
FAO & \textbf{99.93(0.49)} & 0.02(0.09) & 5.25(7.10) & 0.03(0.10) & 0.19(0.80) & 0.06(0.16) & 0.19(0.81) & 0.03(0.11) \\ 
FOM2 & \textbf{100.00(0.00)} & 1.23(0.74) & \textbf{100.00(0.00)} & 2.02(0.94) & 19.60(8.27) & 2.32(0.92) & 59.50(14.36) & 2.02(0.92) \\ 
FAO2 & \textbf{100.00(0.00)} & 1.45(0.86) & \textbf{100.00(0.00)} & 1.56(0.87) & 12.49(7.43) & 2.05(0.97) & 48.37(15.46) & 1.58(0.86) \\ 
ED & \textbf{98.43(2.45)} & 0.00(0.00) & 83.27(7.70) & 0.00(0.00) & 0.03(0.33) & 0.00(0.00) & 0.07(0.47) & 0.00(0.00) \\ 
SEQ1 & \textbf{98.29(2.45)} & 0.00(0.00) & \textbf{100.00(0.00)} & 0.00(0.00) & 0.55(1.47) & 0.00(0.00) & 0.38(1.08) & 0.00(0.00) \\ 
SEQ2 & \textbf{98.29(2.45)} & 1.40(0.70) & \textbf{100.00(0.00)} & 1.13(0.67) & 6.78(6.46) & 0.01(0.05) & 5.39(4.65) & 1.10(0.68) \\ 
SEQ3 & \textbf{98.29(2.45)} & 0.00(0.00) & \textbf{100.00(0.00)} & 0.00(0.00) & 0.06(0.44) & 0.00(0.00) & 0.07(0.47) & 0.00(0.00) \\ 
   \midrule
\multicolumn{9}{c}{$\nu = 5$}\\
 \midrule
\textit{FST} & \textbf{70.97(10.80)} & 11.17(1.68) & 11.85(6.49) & 9.80(1.62) & 42.21(9.51) & 5.77(1.48) & 14.59(6.90) & 10.49(1.69) \\ 
\textit{FSTMG} & 66.31(11.31) & 0.41(0.42) & 1.88(2.54) & 1.69(0.89) & 2.86(3.14) & 1.59(0.92) & 2.13(2.79) & 1.76(0.89) \\ 
\textit{FSTSHA} & 10.29(5.33) & 10.15(1.68) & 9.61(6.09) & 7.97(1.51) & 41.25(9.54) & 3.55(1.10) & 12.10(6.53) & 8.64(1.61) \\
\textit{FSTAM} & 1.72(2.31) & 1.66(0.88) & 2.47(2.77) & 1.64(0.91) & 5.55(4.64) & 1.31(0.78) & 2.91(3.18) & 1.65(0.92) \\
\textit{SF} & \textbf{76.50(9.30)} & 10.47(1.67) & 10.51(6.24) & 9.05(1.67) & 38.99(9.89) & 4.62(1.30) & 13.43(6.69) & 9.74(1.70) \\ 
\textit{SF25} & \textbf{75.65(9.56)} & 10.41(1.64) & 10.39(6.13) & 8.99(1.70) & 38.69(10.57) & 4.65(1.34) & 13.51(6.89) & 9.76(1.72) \\  
\textit{MUOD} & 74.78(18.66) & 14.67(7.06) & 26.25(14.44) & 18.46(10.50) & 44.07(12.90) & 12.76(4.74) & 24.81(12.34) & 17.97(7.90) \\ 
OGMBD & 18.68(9.64) & 4.72(1.40) & 11.17(6.43) & 3.50(1.15) & 26.32(8.14) & 3.35(1.17) & 19.69(8.67) & 2.83(1.12) \\ 
MSPLT & 9.15(6.47) & 2.77(1.27) & 48.01(14.93) & 2.89(1.27) & 27.09(9.30) & 2.80(1.31) & 21.75(10.97) & 2.65(1.25) \\ 
TVD & 18.51(9.74) & 0.00(0.02) & \textbf{100.00(0.00)} & 0.00(0.02) & 0.03(0.30) & 0.00(0.02) & 0.33(1.09) & 0.00(0.02) \\ 
FOM &  \textbf{71.67(12.59)} & 0.07(0.17) & 0.30(1.00) & 0.15(0.26) & 0.67(1.59) & 0.41(0.43) & 0.56(1.46) & 0.43(0.48) \\ 
FAO & 49.99(16.10) & 0.03(0.10) & 0.15(0.76) & 0.07(0.16) & 0.28(1.02) & 0.17(0.26) & 0.28(0.97) & 0.17(0.29) \\ 
FOM2 & \textbf{93.87(4.73)} & 1.25(0.70) & \textbf{100.00(0.00)} & 2.14(0.93) & 8.59(5.64) & 3.14(1.18) & 10.33(6.27) & 2.56(1.02) \\ 
FAO2 & \textbf{93.01(5.61)} & 1.51(0.83) & \textbf{100.00(0.00)} & 1.70(0.91) & 7.07(5.10) & 2.74(1.14) & 8.21(5.57) & 2.09(0.98) \\ 
ED & 16.40(8.54) & 0.00(0.02) & 3.55(3.47) & 0.00(0.02) & 0.00(0.00) & 0.00(0.00) & 0.00(0.00) & 0.00(0.00) \\ 
SEQ1 & 15.25(8.06) & 0.00(0.02) & \textbf{100.00(0.00)} & 0.00(0.02) & 0.01(0.15) & 0.00(0.00) & 0.00(0.00) & 0.00(0.02) \\ 
SEQ2 & 15.68(8.17) & 0.47(0.52) & 60.53(10.52) & 0.22(0.36) & 0.14(0.67) & 0.04(0.12) & 0.13(0.76) & 0.28(0.37) \\ 
SEQ3 & 15.25(8.06) & 0.00(0.02) & \textbf{100.00(0.00)} & 0.00(0.02) & 0.00(0.00) & 0.00(0.00) & 0.00(0.00) & 0.00(0.02) \\ 
\bottomrule
\end{tabular}}
\end{table*}

\end{document}